\documentclass[reprint,nofootinbib,onecolumn,preprintnumbers]{revtex4}
\usepackage{amssymb}
\usepackage{color}

\usepackage{amsthm}
\usepackage{graphicx}
\usepackage{dcolumn}
\usepackage{bm}
\usepackage{subfigure}
\usepackage{amssymb}
\usepackage{verbatim}
\usepackage{amscd}
\usepackage{amssymb}
\usepackage{amsmath, amsfonts}
\usepackage{setspace}
\usepackage[]{graphicx}        
\usepackage{amsthm}
\usepackage{enumerate}
\theoremstyle{plain}
\newtheorem{theorem}{Theorem}
\theoremstyle{plain}

\theoremstyle{plain}
 
\theoremstyle{plain}

\theoremstyle{result}

\theoremstyle{remark}

\theoremstyle{conjecture}

\theoremstyle{observation}

\theoremstyle{definition}

\theoremstyle{corollary}

\theoremstyle{definition}

\theoremstyle{definition}

\theoremstyle{assumption}

\theoremstyle{definition}

\theoremstyle{problem}

\theoremstyle{fact}

\begin{document}

\title{Gapped boundaries, group cohomology and fault-tolerant logical gates}
\author{Beni Yoshida}
\affiliation{Walter Burke Institute for Theoretical Physics, California Institute of Technology, Pasadena, California 91125, USA
}
\affiliation{Perimeter Institute for Theoretical Physics, Waterloo, Ontario  N2L 2Y5, Canada
}
\affiliation{Kavli Institute for Theoretical Physics, University of California, Santa Barbara, CA 93106, USA}

\preprint{NSF-KITP-15-096}

\date{\today}

\begin{abstract}
This paper attempts to establish the connection among classifications of gapped boundaries in topological phases of matter, bosonic symmetry-protected topological (SPT) phases and fault-tolerantly implementable logical gates in quantum error-correcting codes. We begin by presenting constructions of gapped boundaries for the $d$-dimensional quantum double model by using $d$-cocycles functions ($d\geq 2$). We point out that the system supports $m$-dimensional excitations ($m<d$), which we shall call fluctuating charges, that are superpositions of point-like electric charges characterized by $m$-dimensional bosonic SPT wavefunctions. There exist gapped boundaries where electric charges or magnetic fluxes may not condense by themselves, but may condense only when accompanied by fluctuating charges. Magnetic fluxes and codimension-$2$ fluctuating charges exhibit non-trivial multi-excitation braiding statistics, involving more than two excitations. The statistical angle can be computed by taking slant products of underlying cocycle functions sequentially. We find that excitations that may condense into a gapped boundary can be characterized by trivial multi-excitation braiding statistics, generalizing the notion of the Lagrangian subgroup. As an application, we construct fault-tolerantly implementable logical gates for the $d$-dimensional quantum double model by using $d$-cocycle functions. Namely, corresponding logical gates belong to the $d$th level of the Clifford hierarchy, but are outside of the $(d-1)$th level, if cocycle functions have non-trivial sequences of slant products.
\end{abstract}

\maketitle


\section{Introduction}

In studies of theoretical physics, seemingly unrelated subjects of researches have been often found closely connected. In reference~\cite{Beni15}, it has been suggested that classifications of the following three subjects are closely related:
\begin{itemize}
\item gapped boundaries and domain walls in topological phases of matter. 
\item bosonic symmetry-protected topological phases.
\item fault-tolerant logical gates that can be implemented by finite-depth quantum circuits in topological quantum codes.
\end{itemize}
This paper is an attempt to further establish the connection among them. Below, we illustrate the key ideas and summarize the main results of this paper.

Relations between the bulk and the boundary have been important for the understanding of quantum many-body systems. It is known that gapped boundaries in two-dimensional topological phases can be classified by sets of anyons that may condense into gapped boundaries~\cite{Bravyi98, Bombin08, Bais09, Bombin10, Beigi11, Kapustin11, Kitaev12, Levin12, Levin13, Barkeshli13a, Barkeshli13, Fuchs14, Kong14, Lan15, Hung15}. By ``an anyon condenses into a gapped boundary'', we mean that an anyonic excitation can be created and absorbed at the gapped boundary without involving any other excitations. The guiding principle for finding such sets of condensing anyons is to look for the so-called Lagrangian subgroup of anyons which possess (a) mutually-trivial braiding statistics, (b) trivial self-statistics and (c) anyons outside the set have non-trivial mutual statistics with at least one anyon in the set~\cite{Levin13}. While there have been significant progresses toward classification of gapped boundaries supported in two-dimensional topological phases, gapped boundaries in three and higher dimensions are poorly understood except for a certain special family of models~\cite{Walker11, Keyserlingk13}. Based on the success in two dimensions, one might naively conclude that gapped boundaries in higher dimensions can be also classified by the Lagrangian subgroup. It is, however, known that the boundary properties are much richer in higher dimensions; namely, there exist gapped boundaries where excitations may condense only when superpositions of anyonic excitations are involved. The simplest realization of this exotic phenomenon is a certain gapped boundary in a three-dimensional $\mathbb{Z}_{2}\otimes \mathbb{Z}_{2}\otimes \mathbb{Z}_{2}$ topological order (\emph{i.e.} three decoupled copies of the toric code, which is also known as the topological color code in quantum information community~\cite{Bombin06, Bombin07, Kubica15b}) where only the composites of loop-like magnetic fluxes and loop-like superpositions of electric charges may condense~\cite{Beni15, Beni15b}. 

Currently a generic theoretical framework to classify gapped boundaries in higher-dimensional topological phases of matter is missing, and the necessary step would be to explore more examples. We will begin by presenting construction of gapped boundaries for the $d$-dimensional quantum double model with arbitrary finite group $G$ by using $d$-cocycle functions. This generalizes construction of gapped boundaries for the two-dimensional quantum double model by Beigi \emph{et al} to higher dimensions~\cite{Beigi11}. Our construction borrows ideas from recent developments on studies of bosonic symmetry-protected topological (SPT) phases~\cite{Dijkgraaf90, Propitius95, Pollmann12, Chen11, Chen11b, Schuch11, Levin12, Hu13, Chen14, J_Wang15b, J_Wang15c}. Formally, the system with SPT order has certain global on-site symmetry $G$ and its non-degenerate ground state does not break any of the symmetries. A particularly interesting property is the duality between bosonic SPT phases (in the absence of time-reversal symmetry) and intrinsic topological phases via gauging, a process of minimally coupling a system with global symmetry $G$ to gauge fields with gauge symmetry $G$~\cite{Levin12, Hu13, Buerschaper14, Haegeman15}. Under this duality map, a $d$-dimensional bosonic SPT wavefunction, characterized by a $(d+1)$-cocycle function $\nu_{d+1}(g_{0},g_{1},\ldots,g_{d+1})$, is mapped to a ground state of the twisted quantum double model whose braiding statistics is modified from the untwisted one according to the cocycle function $\nu_{d+1}$. In this paper, we will give a precise definition of gauging by formulating it as a bijective and isometric map (\emph{i.e.} a duality map) between wavefunctions with global symmetry and wavefunctions with gauge symmetry. Our construction of gapped boundaries becomes clear by using the gauging map and the duality between intrinsic topological order and SPT order. 

Having constructed gapped boundaries in the quantum double model, we will proceed to characterization of excitations and condensations. For this problem, we shall assume that $G$ is an abelian group for simplicity of discussion. It is well known that the $d$-dimensional quantum double model supports point-like electric charges and codimension-$2$ magnetic fluxes. What may be less familiar is that the $d$-dimensional quantum double model supports $m$-dimensional fluctuating charges ($m \leq d-1$) that are superpositions of electric charges forming $m$-dimensional objects~\cite{Beni15, Beni15b}. We shall demonstrate that $m$-dimensional fluctuating charges can be characterized by $m$-dimensional bosonic SPT wavefunctions based on $(m+1)$-cocycle functions. Such an $m$-dimensional fluctuating charge, living on a boundary $\partial R$ of an $(m+1)$-dimensional region $R$, can be created by a local quantum circuit acting on $R$ while it cannot be created by any local quantum circuit acting exclusively on $\partial R$. In this sense, these are genuine $m$-dimensional objects and are locally unbreakable. We shall see that the special cases for $m=0$ (\emph{i.e.} $0$-dimensional SPT phases) correspond to ordinary point-like electric charges. The remaining task is to find the sets of excitations that may condense into a gapped boundary. For this purpose, it is convenient to introduce the notion of slant products. Formally, a slant product $i_{g}$ for $g\in G$ is a certain map from an $m$-cocycle function $\nu_{m}$ to an $(m-1)$-cocycle function $\nu_{m}^{(g)}:=i_{g}\nu_{m}$. Note that the output of the slant product $i_{g}$ depends on $g\in G$. We will see that composites of a $g$-type magnetic flux and a codimension-$2$ fluctuating electric charge associated with $\nu_{d}^{(g)}$ may condense into a  $\nu_{d}$-type gapped boundary in the $d$-dimensional quantum double model.

We then ask what is the guiding principle for classifying sets of condensing excitations in higher-dimensional topological phases of matter. Recently Wang and Levin have suggested that, in order to fully characterize three-dimensional topological phases of matter, three-loop braiding processes, as well as ordinary loop-loop and loop-particle braiding processes, need to be examined~\cite{Wang14, Wang15}. Inspired by this proposal, we study the three-loop braiding processes among a loop-like fluctuating charge and a pair of loop-like magnetic fluxes in the three-dimensional quantum double model. When a loop-like $\nu_{2}$-charge and a $g_{1}$-type magnetic flux are braided in the presence of a $g_{2}$-type magnetic flux, the resulting statistical angle is given by $\nu_{2}^{(g_{1},g_{2})}:= i_{g_{2}}i_{g_{1}}\nu_{2}\in U(1)$ where slant products are taken sequentially. We then demonstrate that condensing excitations exhibit trivial three-loop braiding statistics. These findings seem to suggest that sets of condensing excitations in three-dimensional topological phases of matter can be classified by a subgroup of excitations (which include both ordinary anyonic excitations and fluctuating charges) that exhibit trivial braiding statistics, including three-loop braiding processes. Our characterization also extends to the quantum double model with $d>3$ where braiding processes involve $d$ different types of codimension-$2$ excitations . 

Finally, we shall apply the aforementioned results to classification of fault-tolerantly implementable logical gates (which is the author's original motivation behind this paper). Given topological quantum error-correcting codes, a long-standing open problem is to find all the logical gates that may be implemented by local quantum circuits, such as transversal implementations of unitary gates~\cite{Gottesman99, Bombin06, Bombin07, Eastin09, Bravyi13b, Brell15, Beverland14, Pastawski15, Kubica15, Kubica15b, Bombin15}. The hope of developing a systematic approach has been elusive so far, and solutions are known only for a few specific examples. 
Recently Bravyi and K\"{o}nig have found that fault-tolerantly implementable logical gates in $d$-dimensional topological stabilizer codes must belong to the $d$th level of the Clifford hierarchy~\cite{Bravyi13b}. Characterization of fluctuating charges and multi-excitation braiding statistics through the framework of group cohomology has interesting applications in constructing fault-tolerantly implementable logical gates. Consider a $(d-1)$-dimensional fluctuating charge and sweep it over the entire $d$-dimensional quantum double model. Since this process changes types of excitations, it implements some non-trivial action on the ground state space of the quantum double model, and thus is a fault-tolerant logical operation. Recall that the notion of the Clifford hierarchy has natural generalization for any abelian finite group $G$~\cite{Gottesman99, Watson15}. We find that a logical unitary, corresponding to a $\nu_{d}$-charge, is a non-trivial $d$th level logical gate if there exist $g_{1},\ldots,g_{d}\in G$ such that the sequential slant product $\nu_{d}^{(g_{1}\ldots g_{d})}:=i_{g_{d}}i_{g_{d-1}}\ldots i_{g_{1}}\nu_{d}$ is non-trivial. This surprising connection between group cohomology and the Clifford hierarchy can be understood from the following intuition. Multi-excitation braiding processes can be expressed as sequential group commutators among codimension-$1$ world-sheet operators that characterize propagations of codimension-$2$ excitations. At the same time, sequential group commutators naturally appear in a definition of the Clifford hierarchy~\cite{Pastawski15}. Our finding will allow one to find a large number of non-trivial logical gates, belonging to the higher-level of the generalized Clifford hierarchy in a systematic manner, shedding a new light on theories of fault-tolerant quantum computation. We believe that our technique is also applicable to the twisted quantum double model with non-abelian $G$. 

In summary, we shall present a systematic framework of constructing gapped boundaries and fault-tolerantly implementable logical gates in the quantum double model by borrowing the idea of the gauging map utilized in studies of bosonic SPT phases. Along the way, we will find new types of gapped boundaries that do not fit into conventional classification scheme of gapped boundaries of topological phases. We will also see various examples of non-trivial multi-excitation braiding processes, such as three-loop braiding. The rest of the paper is dedicated to detailed descriptions of these ideas and findings. 

\section{Gapped boundary for higher-dimensional quantum double model}

In this section, we present constructions of gapped boundaries for the $d$-dimensional quantum double model by using $d$-cocycle functions. We begin by defining the procedure of \emph{gauging} as a duality map (a bijective and isometric map) from wavefunctions with global symmetries to wavefunctions with gauge symmetries. We then construct gapped boundaries by using bosonic SPT wavefunctions which are built on the machinery of group cohomology. This generalizes constructions of gapped boundaries by Beigi \emph{et al} to higher-dimensional systems ($d>2$). 

\subsection{Quantum double model}

\begin{figure}[htb!]
\centering
\includegraphics[width=0.55\linewidth]{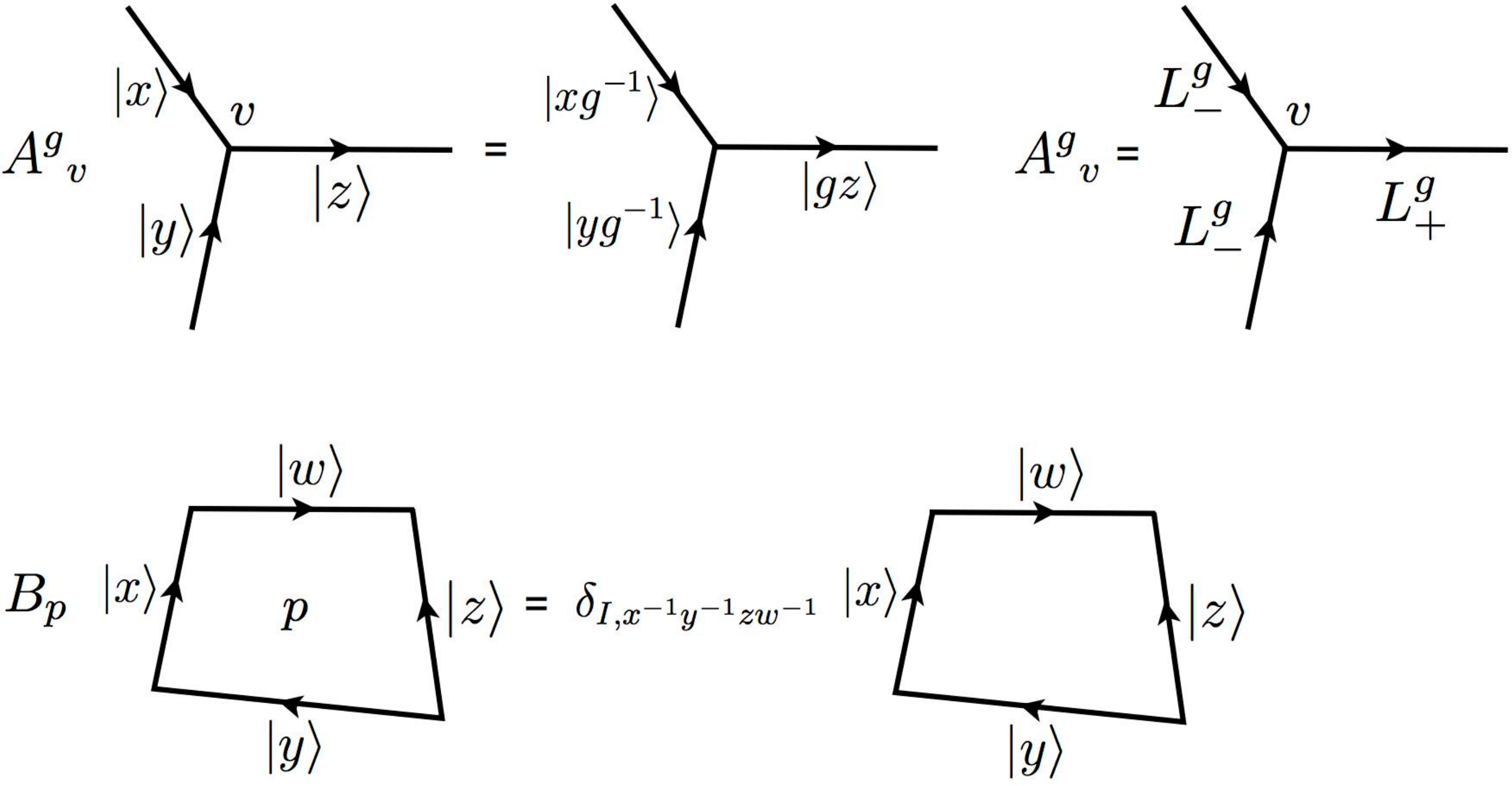}
\caption{Definitions of operators ${A^{g}}_{v}$ and $B_{p}$. Here $I$ represents an identity element in $G$, $L^{g}_{+}|h\rangle =|gh\rangle$ and $L^{g}_{-}|h\rangle =|hg^{-1}\rangle$.
} 
\label{fig_operator}
\end{figure}

Let us begin by recalling construction of the quantum double model in $d$-spatial dimensions~\cite{Kitaev03, Wan15}. Consider a $d$-dimensional lattice $\Lambda$ with directed edges. A Hilbert space with the orthogonal basis $\{ |g\rangle : g\in G \}$ is associated to each edge where $G$ is a finite group. The entire Hilbert space is denoted by $\mathcal{H}_{1}$. Operators ${A^{g}}_{v}$ and $B_{p}$ are defined according to Fig.~\ref{fig_operator} where $B_{p}$ is a projector onto a subspace with no flux on a plaquette $p$. Define
\begin{align}
A_{v} = \frac{1}{|G|} \sum_{g\in G} {A^{g}}_{v}.
\end{align}
Operators $A_{v}$ and $B_{p}$ are projectors and pairwise commute. The Hamiltonian of the $d$-dimensional quantum double model is given by 
\begin{align}
H_{G}= - \sum_{v} A_{v} - \sum_{p} B_{p}
\end{align}
where the summations run over all vertices $v$ and plaquettes $p$. The ground state $|\psi_{gs}\rangle$ satisfies 
\begin{align}
A_{v}|\psi_{gs}\rangle=B_{p}|\psi_{gs}\rangle=|\psi_{gs}\rangle
\end{align}
for all $v,p$. For an arbitrary closed loop $\gamma$, one can also define a projection operator $B(\gamma)$ onto fluxless subspaces. We then have $B(\gamma)|\psi_{gs}\rangle=|\psi_{gs}\rangle$ for any contractible loop $\gamma$. 

\subsection{Gauging map}

\begin{figure}[htb!]
\centering
\includegraphics[width=0.45\linewidth]{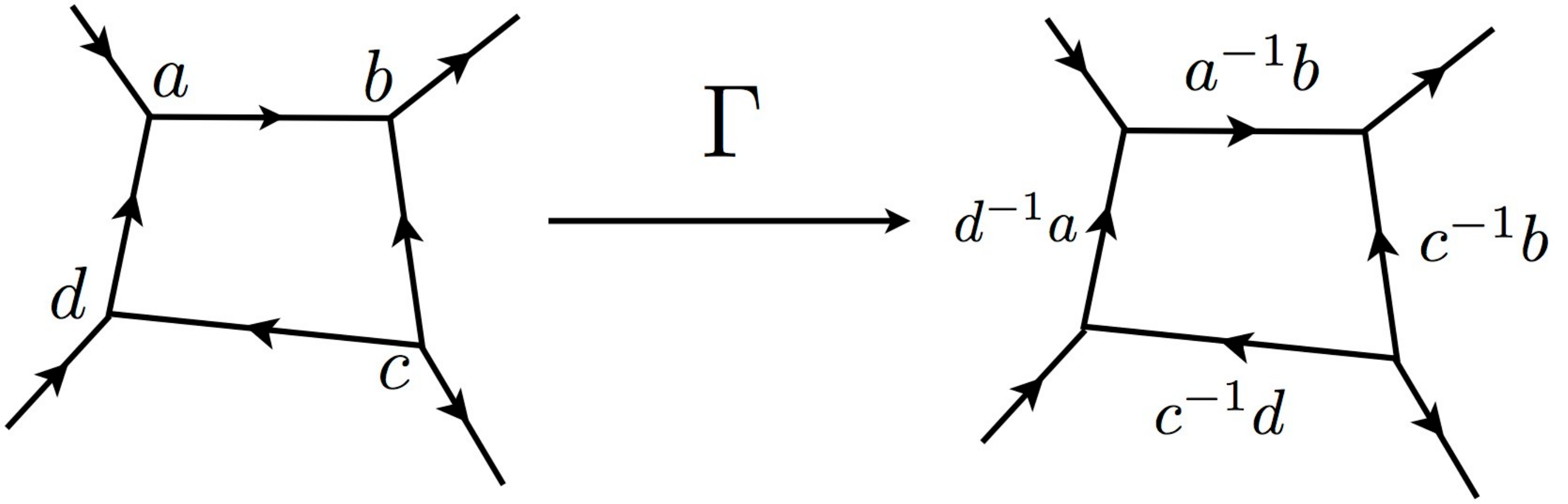}
\caption{The gauging map. The output wavefunction is given by taking group multiplication. The output wavefunctions always satisfy gauge constraints. 
} 
\label{fig_map}
\end{figure}

Next let us present a precise definition of gauging. By now, gauging has become a standard technique for studying SPT phases. Yet it would be beneficial to formulate the procedure of gauging in a rigorous manner. Formally, the gauging is an isometric bijective map (\emph{i.e.} a duality map) between wavefunctions with global symmetry $G$ to wavefunctions with gauge symmetry $G$. Consider the same lattice $\Lambda$ as before, but with the Hilbert space associated to vertices instead of edges. The entire Hilbert space is denoted by $\mathcal{H}_{0}$. Here the subscript $0$ indicates that spins live on $0$-dimensional objects. The gauging map $\Gamma$ between computational basis states in $\mathcal{H}_{0}$ and $\mathcal{H}_{1}$ is defined according to Fig.~\ref{fig_map} by taking group multiplications: 
\begin{align}
\Gamma : |v_{1},v_{2},\ldots, v_{n}\rangle  \rightarrow |e_{1},e_{2},\ldots, e_{m}\rangle \qquad v_{i},e_{j}\in G
\end{align}
where $n,m$ are the total numbers of vertices and edges on $\Lambda$. Let us briefly summarize key properties of the gauging map (assuming that $\Lambda$ is a connected graph).

\begin{itemize}
\item The output state $|\hat{\psi}\rangle = \Gamma(|\psi\rangle) \in \mathcal{H}_{1}$ always satisfies 
\begin{align}
B(\gamma)|\hat{\psi}\rangle=|\hat{\psi}\rangle
\end{align}
for an arbitrary closed loop $\gamma$, including non-contractible loops. 
\item Output wavefunctions are identical, 
\begin{align}
\Gamma(|v_{1}',v_{2}',\ldots, v_{n}'\rangle ) = \Gamma(|v_{1},v_{2},\ldots, v_{n}\rangle )
\end{align}
if and only if $\exists g\in G$ such that $(v_{1},v_{2},\ldots,v_{n})=(gv_{1}',gv_{2}',\ldots,gv_{n}')$.
\end{itemize}

The map $\Gamma$ can be extended to wavefunctions which are superpositions of computational basis states. Consider a symmetric subspace of $\mathcal{H}_{0}$ defined by 
\begin{align}
\mathcal{H}_{0}^{sym} = \{ |\psi\rangle \in \mathcal{H}_{0} : S^{g}|\psi\rangle = |\psi\rangle \ \ \forall g\in G \}.
\end{align}
Here $L_{+}^{g}|h\rangle:=|gh\rangle$ for $g,h\in G$, and global symmetry operators are 
\begin{align}
S^{g}:=\bigotimes_{v} {L^{g}_{+}}_{v}. 
\end{align}
Also, consider a gauge symmetric subspace of $\mathcal{H}_{1}$ defined by 
\begin{align}
\mathcal{H}_{1}^{sym} = \{ |\hat{\psi}\rangle \in \mathcal{H}_{1} : B(\gamma) |\hat{\psi}\rangle = |\hat{\psi}\rangle \ \ \forall \gamma \}.
\end{align}
Here $\gamma$ is an arbitrary closed loop, either contractible or non-contractible. We consider the gauging map $\Gamma$ such that $\Gamma : \mathcal{H}_{0}^{sym} \rightarrow \mathcal{H}_{1}^{sym}$. More explicitly, for $|\psi\rangle = \sum_{v_{1},v_{2},\ldots} C_{v_{1},v_{2},\ldots}|v_{1},v_{2},\ldots \rangle\in \mathcal{H}_{0}^{sym}$, the gauging map $\Gamma$ is given by
\begin{align}
\Gamma (|\psi\rangle) = \frac{1}{\sqrt{|G|}} \sum_{v_{1},v_{2},\ldots} C_{v_{1},v_{2},\ldots} \Gamma(|v_{1},v_{2},\ldots \rangle)\in \mathcal{H}_{1}^{sym}.
\end{align}
Here the normalization factor comes from the fact that there are $|G|$ computational basis states with the same output. 

\begin{itemize}
\item Note that $\dim \mathcal{H}_{0}^{sym} = \dim \mathcal{H}_{1}^{sym}$. Moreover, the gauging map $\Gamma$ is a one-to-one map, and is an isometry, meaning that it preserves the inner products of input states and output states: 
\begin{align}
\langle \psi |\psi'\rangle = \langle \hat{\psi} |\hat{\psi'}\rangle \qquad |\hat{\psi}\rangle,|\hat{\psi'}\rangle = \Gamma(|\psi\rangle),\Gamma(|\psi'\rangle). 
\end{align}
In other words, the gauging map $\Gamma : \mathcal{H}_{0}^{sym}\rightarrow \mathcal{H}_{1}^{sym}$ is a duality map.
\end{itemize}

The gauging map $\Gamma$ transforms a trivial symmetric wavefunction into a fluxless ground state of the quantum double model. Consider an arbitrary symmetric state $|\psi\rangle \in \mathcal{H}_{0}^{sym}$ and its output state $|\hat{\psi}\rangle=\Gamma(|\psi\rangle)$. Observe that applications of ${L^{g}_{-}}_{v}$ on $\mathcal{H}_{0}^{sym}$ is equivalent to applications of ${A^{g}}_{v}$ on $\mathcal{H}_{1}$:
\begin{align}
{A^{g}}_{v}\Gamma(|\psi\rangle)=\Gamma({L^{g}_{-}}_{v}|\psi\rangle)\qquad \forall|\psi\rangle \in \mathcal{H}^{sym}_{0}.
\end{align}
Consider a symmetric product state $|+\rangle^{\otimes n}$ where $|+\rangle := \frac{1}{\sqrt{|G|}}\sum_{g\in G} |g\rangle$. Note ${L^{g}_{-}}_{v}|+\rangle^{\otimes n}=|+\rangle^{\otimes n}$. Then the output state $|\psi_{gs}\rangle = \Gamma(|+\rangle^{\otimes n})$ must be a ground state of the quantum double Hamiltonian $H_{G}$ since $A_{v}|\psi_{gs}\rangle=B_{p}|\psi_{gs}\rangle=|\psi_{gs}\rangle$. More generically, let $\mathcal{V}_{0},\mathcal{V}_{1}$ be sets of unitary operators which preserve $\mathcal{H}_{0}^{sym},\mathcal{H}_{1}^{sym}$ respectively. Given $U\in \mathcal{V}_{0}$, there always exists a corresponding operator $\hat{U}\in  \mathcal{V}_{1}$ which satisfies the following\footnote{The gauging map is \emph{not} a physical process that can be implemented by any local quantum circuit plus adding/removing ancilla since topologically ordered states are generated from trivial states. Yet, the gauging map preserves the geometric locality of unitary operators as long as they are supported on a connected region. However the gauging map does not preserve the locality of unitary operators when they are supported on non-connected regions. For instance, if an operator is supported on two spatially separated spheres, then the corresponding operators may require supports on line-like regions connecting two spheres.}: 
\begin{align}
\Gamma(U |\psi\rangle)=\hat{U} \Gamma (|\psi\rangle) \qquad \forall |\psi\rangle \in \mathcal{H}_{0}^{sym}. \label{eq:correspondence}
\end{align}
Note that corresponding operators are not unique. For instance, if $\hat{U}$ is a corresponding operator of $U$, then $\hat{U}B(\gamma)$ for any closed loop $\gamma$ is also a corresponding operator since $B(\gamma)$ acts as an identity inside $\mathcal{H}_{1}^{sym}$. 

It is worth looking at an example. Let $G=\mathbb{Z}_{2}$ and $d=2$. Consider a square lattice on a torus where qubits live on vertices. The trivial Hamiltonian with global $\mathbb{Z}_{2}$ on-site symmetry is given by $H = - \sum_{v} X_{v}$, and the ground state is $|\psi\rangle = |+\rangle^{\otimes n}$ where $|+\rangle=\frac{1}{\sqrt{2}}(|0\rangle + |1\rangle)$. The Hamiltonian and its ground state are symmetric under $\mathbb{Z}_{2}$ global transformation: $S = \bigotimes_{v} X_{v}$ since $S|\psi\rangle = |\psi\rangle$ and $SHS^{\dagger}=H$. Let $|\hat{\psi}\rangle$ be the output wavefunction of the gauging map. Let us see that $|\hat{\psi}\rangle$ is a ground state of the two-dimensional toric code: $\hat{H} = -\sum_{v} A_{v} - \sum_{p} B_{p}$. Indeed, gauge constraints account for the plaquette terms $B_{p}=\otimes_{e\in p} Z_{e}$ in the toric code. As for the star terms $A_{v}=\otimes_{e\in v} X_{e}$, observe that $X_{v}|\psi\rangle=|\psi\rangle$ for a symmetric system. Flipping a spin at vertex $v$ is equivalent to flipping four spins connected to the vertex in the gauge theory. Thus, $X_{v}$ operator in the symmetric system corresponds to the vertex term $A_{v}$ in the gauge theory: $\hat{X}_{v}= A_{v}$. As such, the output wavefunction satisfies $A_{v}|\hat{\psi}\rangle=|\hat{\psi}\rangle$ for all $v$.

\subsection{Bosonic SPT phases}

\begin{figure}[htb!]
\centering
\includegraphics[width=0.45\linewidth]{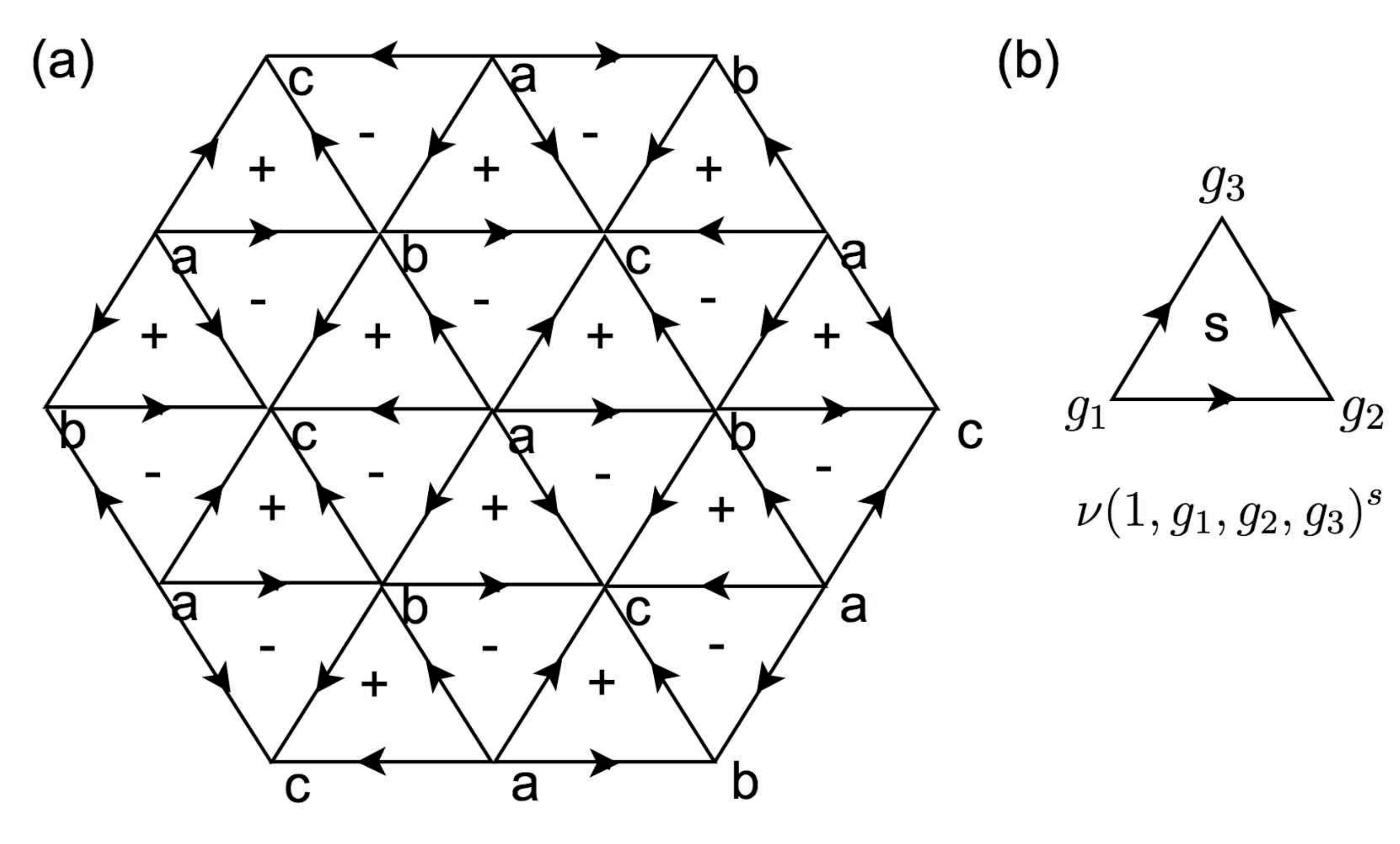}
\caption{(a) A three-colorable graph in two dimensions. (b) A cocycle function where $s=\pm 1$ represents the parity. 
} 
\label{fig_colorable}
\end{figure}

Another important ingredient is the classification of bosonic SPT phases by the machinery of group cohomology which is briefly reviewed in this subsection~\cite{Chen13}. In this paper, we shall consider constructions supported on a closed colorable graph $\Lambda$, a $d$-dimensional graph which is $d+1$ colorable, in a sense that $d+1$ distinct color labels can be assigned to vertices such that neighboring vertices have different colors. An example of a colorable graph is a two-dimensional triangular lattice which is three colorable (Fig.~\ref{fig_colorable}(a)). Colorability of graphs is not essential, but simplifies calculations. 

Let us first recall the basic definition of group cocycles. 

\begin{itemize}
\item A $d$-cochain of group $G$ is a complex function $\nu_{d}(g_{0},g_{1},\ldots,g_{d})$ of $d+1$ variables in $G$ that satisfy 
\begin{align} 
&|\nu_{d}(g_{0},g_{1},\ldots,g_{d})|=1 \\
&\nu_{d}(g_{0},g_{1},\ldots,g_{d})=\nu_{d}(gg_{0},gg_{1},\ldots,gg_{d}), \qquad g \in G.
\end{align}
\item The $d$-cocycles are $d$-cochains that satisfy cocycle conditions
\begin{align}
\prod_{j=0}^{d+1} \nu_{d}^{(-1)^{j}}(g_{0},\ldots, g_{j-1},g_{j+1},\ldots,g_{d+1})=1.
\end{align}
\item The $d$-coboundaries $\lambda_{d}$ are $d$-cocycles that can be constructed from $(d-1)$-cochains $\nu_{d-1}$:
\begin{align}
\lambda_{d}(g_{0},g_{1},\ldots,g_{d})=\prod_{j=0}^{d} \mu_{d-1}^{(-1)^{j}}(g_{0},\ldots, g_{j-1},g_{j+1},\ldots,g_{d+1}).
\end{align}
\item Two $d$-cocycles $\nu_{d}$ and $\nu_{d}'$ are said to be equivalent iff they differ by a coboundary $\nu_{d}=\nu_{d}'\lambda_{d}$.  
\end{itemize}

Fixed-point Hamiltonians and wavefunctions for bosonic SPT phases can be constructed as follows. Let $a_{1},\ldots,a_{d+1}$ be color labels. (In Fig.~\ref{fig_colorable}, we took $a_{1}=a$, $a_{2}=b$ and $a_{3}=c$). A natural choice of orientations for a colorable graph is to draw an arrow from a vertex of color $a_{i}$ to a vertex of color $a_{j}$ if $i<j$. On a colorable graph, one can assign parity $S(\Delta)=\pm1$ to each $d$-simplex $\Delta$ such that neighboring $d$-simplexes have opposite parities~\cite{Kubica15} (see Fig.~\ref{fig_colorable}(a)(b)). A $|G|$-dimensional Hilbert space is associated to each vertex of $\Lambda$. Consider a trivial Hamiltonian 
\begin{align}
H_{trivial} = - \sum_{v} {L_{-}}_{v} \qquad {L_{-}}_{v}=\sum_{g} {L^{g}_{-}}_{v}  
\end{align}
whose ground state is $|\psi_{0}\rangle = |+\rangle^{\otimes n}$. A non-trivial SPT Hamiltonian can be obtained by applying some appropriate local unitary transformation $U_{\nu_{d+1}}$ to a trivial Hamiltonian. 

\begin{itemize}
\item The quantum circuit $U_{\nu_{d+1}}$ is a diagonal phase gate, and can be written as follows:
\begin{align}
U_{\nu_{d+1}} = \prod_{\Delta} U_{\Delta} \qquad 
U_{\Delta}|g_{1}^{\Delta},\ldots,g_{d+1}^{\Delta}\rangle 
=\nu_{d+1}(1,g_{1}^{\Delta},\ldots,g_{d+1}^{\Delta})^{S(\Delta)}|g_{1}^{\Delta},\ldots,g_{d+1}^{\Delta}\rangle
\end{align}
where $\nu_{d+1}(1,g_{1}^{\Delta},\ldots,g_{d+1}^{\Delta})$ is a $(d+1)$-cocycle function, and $g_{j}^{\Delta}$ corresponds to a vertex of color $a_{j}$ in $\Delta$. Here $d$-simplexes are denoted by $\Delta$, and the product runs over all $d$-simplexes. 
\item The nontrivial SPT Hamiltonian and wavefunction are given by
\begin{align}
H_{0}=U_{\nu_{d+1}}H_{trivial}U_{\nu_{d+1}}^{\dagger} \qquad |\psi\rangle=U_{\nu_{d+1}}|\psi_{trivial}\rangle 
\end{align}
where the SPT wavefunction $|\psi\rangle$ is symmetric: $S^{g} |\psi\rangle = |\psi\rangle$.
\item The SPT wavefunction $|\psi\rangle$ is non-trivial, in the sense that there is no finite-depth symmetric quantum circuit to crate the wavefunction, if and only if the cocycle function $\nu_{d+1}$ is non-trivial.
\end{itemize}

As for the second statement, it suffices to show the following relation:
\begin{align}
\prod_{\Delta} \nu_{d+1}^{S(\Delta)}(1,g_{1}^{\Delta},\ldots,g_{d+1}^{\Delta})
= \prod_{\Delta} \nu_{d+1}^{S(\Delta)}(h,g_{1}^{\Delta},\ldots,g_{d+1}^{\Delta})\qquad \forall h\in G. \label{eq:symmetry}
\end{align}
The cocycle conditions imply 
\begin{align}
\nu_{d+1}(h,g_{1},\ldots,g_{d+1})
\nu_{d+1}(1,g_{1},\ldots,g_{d+1})^{-1}
\prod_{j=1}^{d+1}\nu_{d+1}(1,g_{1},\ldots,g_{j-1},h,g_{j+1},\ldots,g_{d+1})^{(-1)^{j+1}}=1.
\end{align}
Given a $(d-1)$-simplex, there always exist a pair of $d$-simplexes, $\Delta,\Delta'$ which neighbor to each other through this $(d-1)$-simplex. Since neighboring $d$-simplexes have opposite parities and $\Lambda$ is a closed graph, one has
\begin{align}
\prod_{\Delta}\prod_{j=1}^{d+1}\nu_{d+1}(1,g^{\Delta}_{1},\ldots,g^{\Delta}_{j-1},h,g^{\Delta}_{j+1},\ldots,g^{\Delta}_{d+1})^{(-1)^{j+1}}=1.
\end{align}
Thus, one has Eq.~(\ref{eq:symmetry}). As for the third statement, let us apply the gauging map with respect to $G$. Then the output wavefunctions $|\hat{\psi}_{trivial}\rangle$ and $|\hat{\psi}\rangle$ are ground states of topologically ordered Hamiltonians with different braiding statistics, which implies that the original symmetric wavefunctions $|\psi_{trivial}\rangle$ and $|\psi\rangle$ belong to different quantum phases in the presence of global symmetries. It is well known that the topological model, obtained via gauging, is identical to the $d$-dimensional Dijkgraaf-Witten topological gauge theory~\cite{Hu13,Wan15}. 

It is worth looking at an example. For $d=1$ and $G=\mathbb{Z}_{2}\otimes \mathbb{Z}_{2}$, a non-trivial $2$-cocycle function is given by
\begin{align}
\nu_{2}(1,g,h)=(-1)^{g_{1}g_{2}}(-1)^{g_{1}h_{2}},\qquad g=(g_{1},g_{2})\quad h=(h_{1},h_{2}).
\end{align}
where $g_{1},g_{2},h_{1},h_{2}=0,1$. This leads to the SPT Hamiltonian with $\mathbb{Z}_{2}\otimes \mathbb{Z}_{2}$ symmetry with the following terms:
\begin{align}
\includegraphics[height=1.3in]{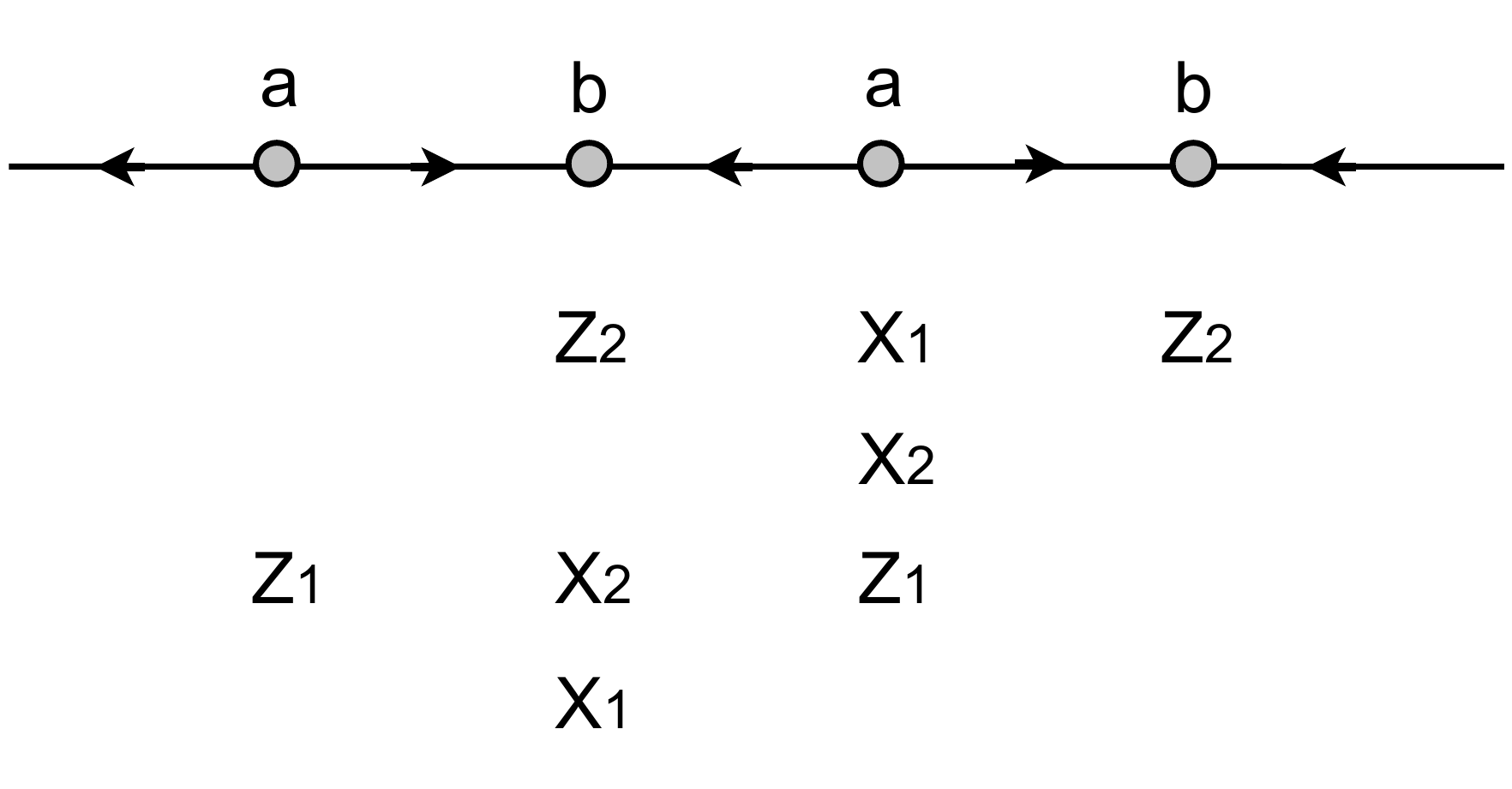}
\end{align}
where $X_{1},Z_{1},X_{2},Z_{2}$ are Pauli operators acting on the first and second qubits at each site. Note that interaction terms depend on colors of vertices. 

\subsection{Gapped boundary}

\begin{figure}[htb!]
\centering
\includegraphics[width=0.35\linewidth]{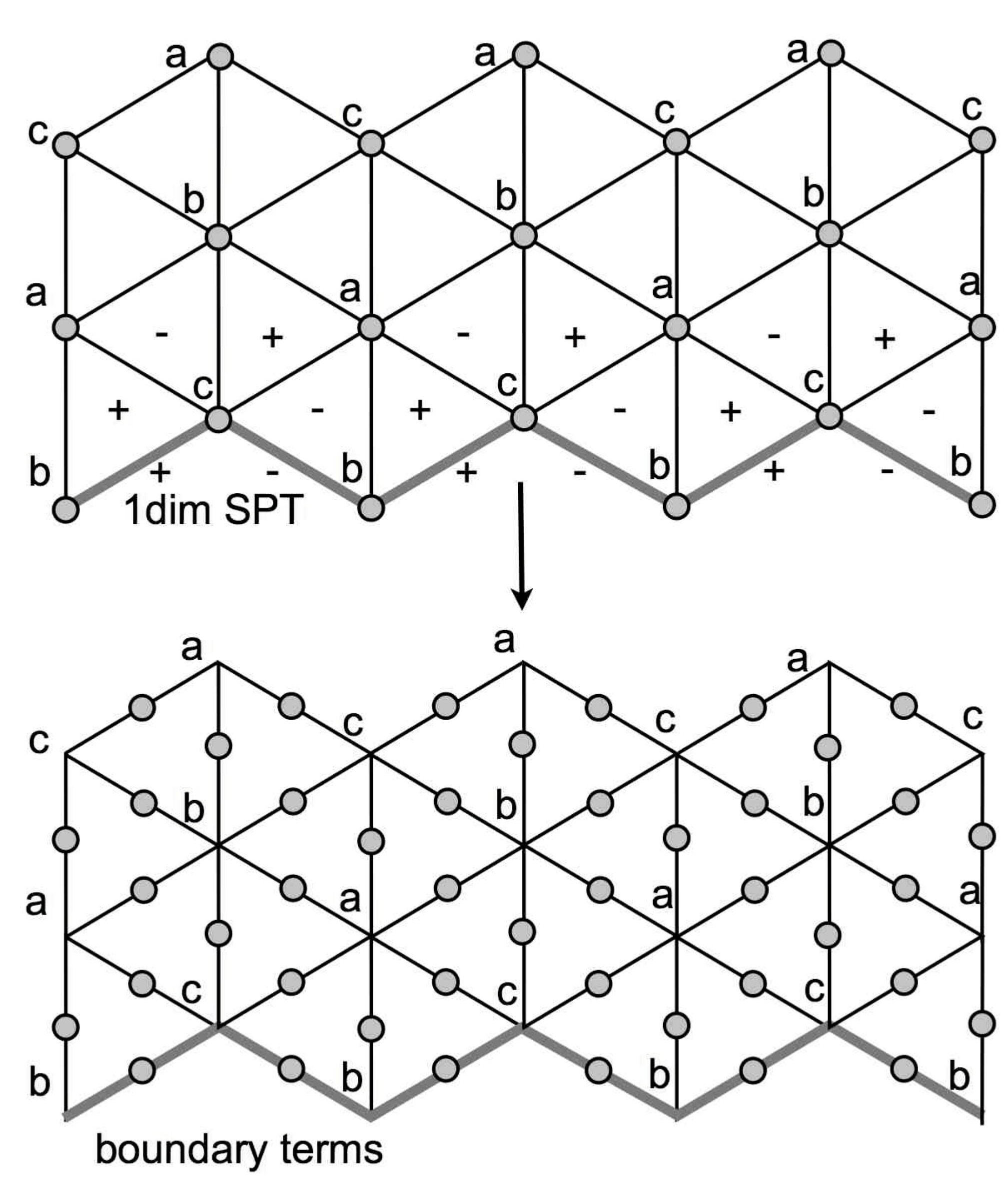}
\caption{Gapped boundary by gauging $(d-1)$-dimensional SPT phases. The figure shows the cases for $d=2$. 
} 
\label{fig_boundary}
\end{figure}

Now we construct gapped boundaries for the $d$-dimensional quantum double model. Consider a $d$-dimensional $(d+1)$-colorable graph $\Lambda$ with color labels $a_{0},a_{1},\ldots,a_{d}$. Assume that the graph has a boundary $\partial \Lambda$ which is $d$-colorable with color labels $a_{1},\ldots,a_{d}$. Parity of each $(d-1)$-simplex in the boundary graph $\partial \Lambda$ is given by $S(\Delta)$ where $\Delta$ is a $d$-simplex containing the $(d-1)$-simplex. Consider the following symmetric wavefunction:
\begin{align}
|\psi\rangle = |\psi_{trivial}\rangle \otimes |\psi_{SPT}(\nu_{d})\rangle
\end{align}
where $|\psi_{trivial}\rangle$ is a trivial $G$-symmetric wavefunction supported on $\Lambda\setminus \partial \Lambda$ and $|\psi_{SPT}(\nu_{d})\rangle$ is a $(d-1)$-dimensional $G$-symmetric SPT wavefunction supported on $\partial \Lambda$, corresponding to some non-trivial $d$-cocycle function $\nu_{d}$. The Hamiltonian for this wavefunction can be written as 
\begin{align}
H = -\sum_{v \in \Lambda\setminus \partial \Lambda } {L_{-}}_{v} - \sum_{v \in \partial \Lambda} O_{v}
\end{align}
where $O_{v}$ are interaction terms for the $(d-1)$-dimensional SPT Hamiltonian. Since the entire wavefunction is $G$-symmetric, the gauging map $\Gamma$ can be applied (Fig.~\ref{fig_boundary}):
\begin{align}
\hat{H} = - \sum_{v \in \Lambda\setminus \partial \Lambda} A_{v} - \sum_{v \in \partial \Lambda} \hat{O}_{v} - \sum_{p} B_{p}
\end{align}
where $\hat{O}_{v}$ represents corresponding operators of $O_{v}$. This Hamiltonian can be viewed as the $d$-dimensional quantum double model with boundary terms $\hat{O}_{v}$. 

\begin{figure}[htb!]
\centering
\includegraphics[width=0.50\linewidth]{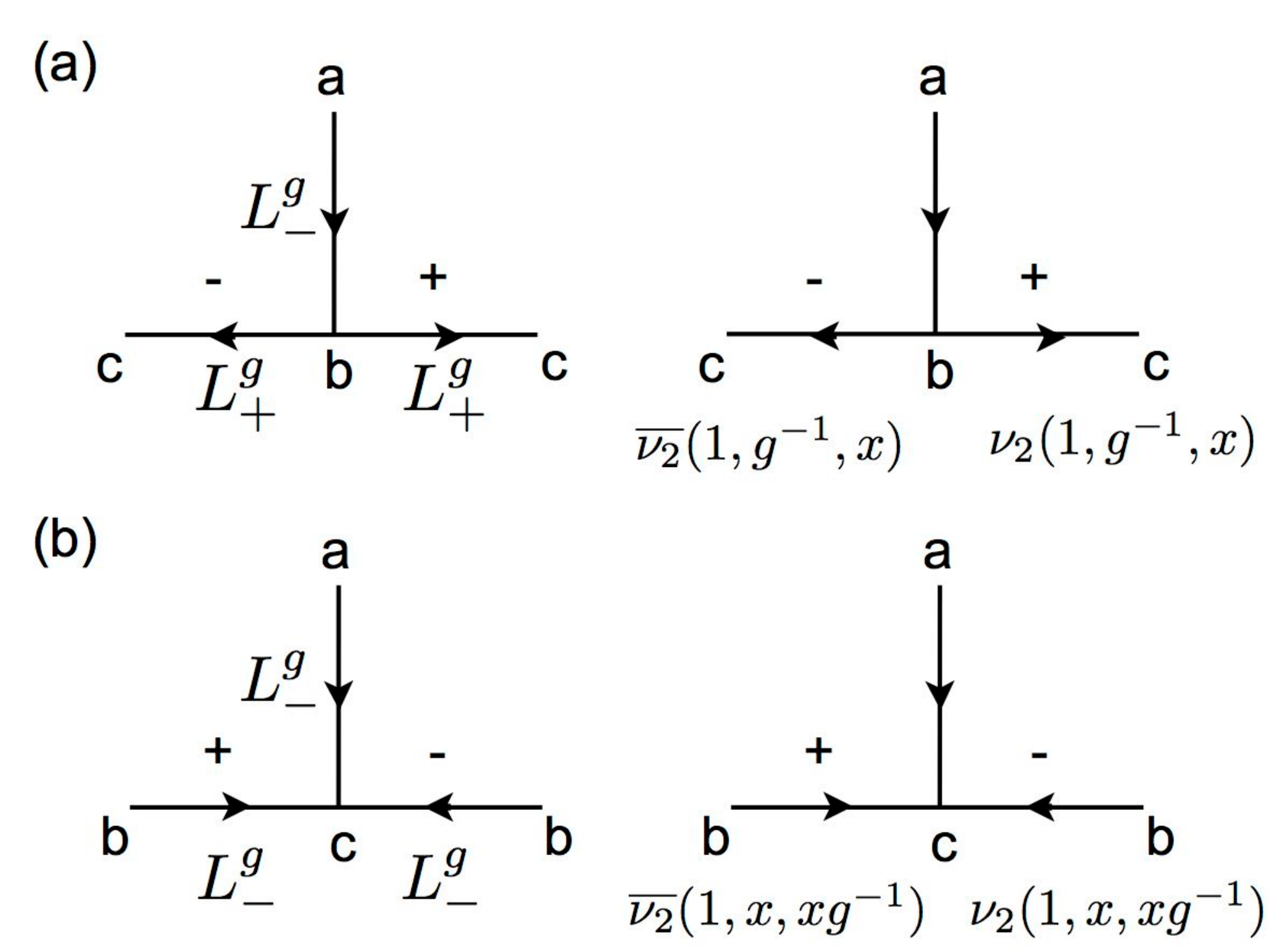}
\caption{Boundary terms in the two-dimensional quantum double model. Here $x$ represents the spin value on the corresponding edge. The order is important. Terms on the right hand side acts first. (a) Terms at vertices of color $b$. (b) Terms at vertices of color $c$. 
} 
\label{fig_2dim_terms}
\end{figure}

\begin{figure}[htb!]
\centering
\includegraphics[width=0.65\linewidth]{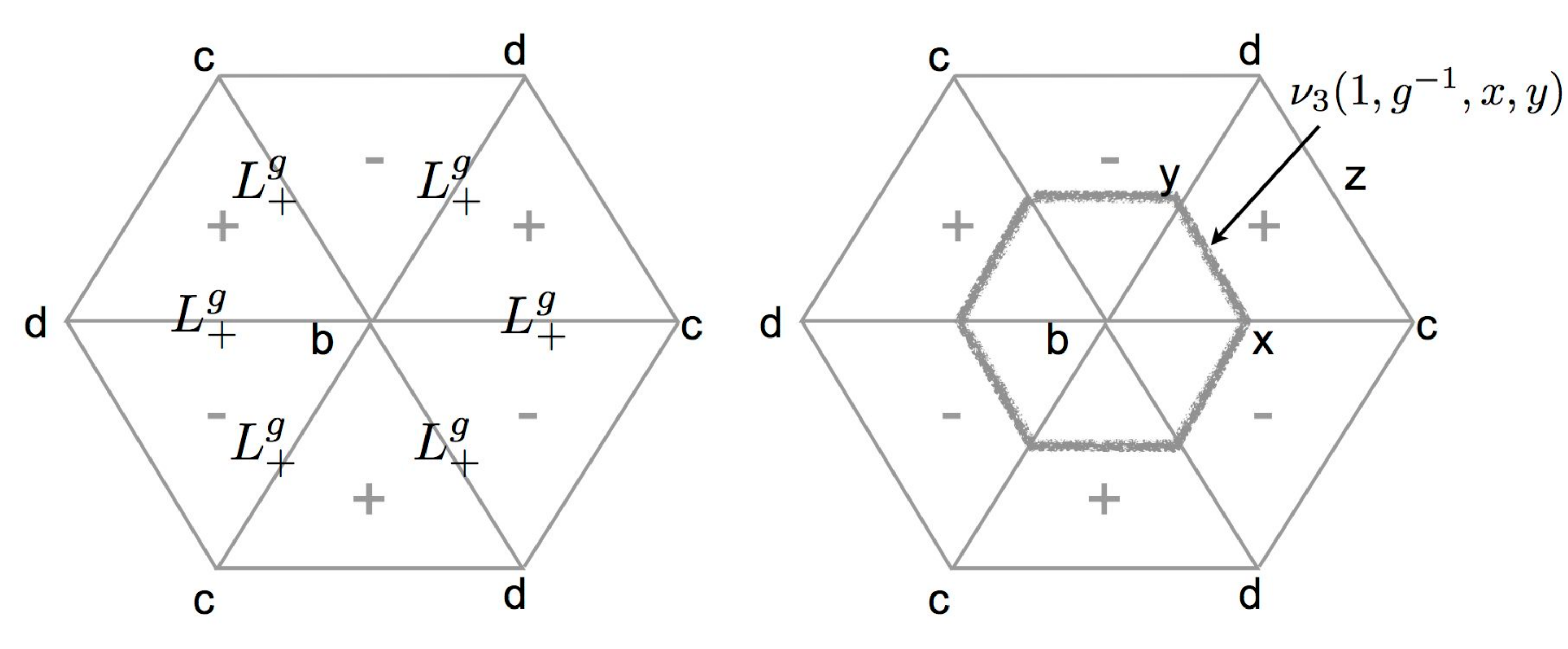}
\caption{Boundary terms in the three-dimensional quantum double model for vertices of color $b$. The vertex term is decorated by a product of two-body phase operators. Again the order is important. 
} 
\label{fig_3dim_terms}
\end{figure}

Explicit forms of $O_{v}$ and $\hat{O}_{v}$ can be found by standard calculations of cocycle functions. Let $U_{\nu_{d}}$ be the encoding circuit for the $(d-1)$-dimensional SPT wavefunction acting on the boundary graph $\partial \Lambda$. Then
\begin{align}
O_{v}=\sum_{g}U_{\nu_{d}}{L^{g}_{-}}_{v}U_{\nu_{d}}^{\dagger} = \sum_{g} {L^{g}_{-}}_{v}K({L^{g}_{-}}_{v},U_{\nu_{d}}^{\dagger})
\end{align}
where $K({L^{g}_{-}}_{v},U_{\nu_{d}}^{\dagger})={{L^{g}_{-}}_{v}}^{\dagger}U_{\nu_{d}}{L^{g}_{-}}_{v}U_{\nu_{d}}^{\dagger}$ is a group commutator. Assume that $v$ is a vertex of color $a_{j}$. Notice that $K({L^{g}_{-}}_{v},U^{\dagger}_{\nu_{d}})$ is a diagonal phase operator. It can be written as a product of $d$-body phase operators acting on $(d-1)$-simplexes containing $v$:
\begin{align}
K({L^{g}_{-}}_{v},U_{\nu_{d}}^{\dagger}) = \prod_{v \in \Delta}T_{\nu_{d}(g_{1}^{\Delta},g_{2}^{\Delta},\ldots, g_{j}^{\Delta}, g_{j}^{\Delta}g^{-1},\ldots, g_{d}^{\Delta} )^{(-1)^{j}\cdot s(\Delta)\cdot (-1)}}
\end{align}
where $\Delta$ represents $(d-1)$-simplexes. This expression can be obtained by using cocycle conditions. Here $T_{\nu(g_{1}^{\Delta},g_{2}^{\Delta},\ldots, g_{j}^{\Delta},g_{j}^{\Delta}g^{-1},\ldots, g_{d}^{\Delta} )}$ represents a phase operator which adds $U(1)$ phase $\nu(g_{1}^{\Delta},g_{2}^{\Delta},\ldots, g_{j}^{\Delta}, g_{j}^{\Delta}g^{-1},\ldots, g_{d}^{\Delta} )$ to a $d$-spin state $|g_{1}^{\Delta},g_{2}^{\Delta},\ldots, g_{d}^{\Delta}\rangle$. The corresponding operator $\hat{O}_{v}$ can be obtained by noticing 
\begin{align}
\nu(g_{1}^{\Delta},g_{2}^{\Delta},\ldots, g_{j}^{\Delta}, g_{j}^{\Delta}g^{-1},\ldots, g_{d}^{\Delta} ) = 
\nu(1,\tilde{g_{2}}^{\Delta},\ldots, \tilde{g_{j}}^{\Delta}, \tilde{g_{j}}^{\Delta}g^{-1},\ldots, \tilde{g_{d}}^{\Delta} )
\end{align}
where $\tilde{g_{i}}=g_{1}^{-1}g_{i}$ corresponds to spin values for edges connecting vertices of color $a_{1}$ and color $a_{i}$ after the gauging map. Thus, the corresponding operator on the boundary is given by
\begin{align}
\hat{O}_{v}= \sum_{g}{A^{g}}_{v}\prod_{v\in \Delta} T_{\nu(1,\tilde{g_{2}}^{\Delta},\ldots, \tilde{g_{j}}^{\Delta},\tilde{g_{j}}^{\Delta}g^{-1},\ldots, \tilde{g_{d}}^{\Delta} )^{(-1)^{j}\cdot s(\Delta)\cdot (-1)}}.
\end{align}
Boundary terms for $d=2$ are graphically shown in Fig.~\ref{fig_2dim_terms}(a)(b) where ordinary vertex terms $A^{g}$ are decorated by phase operators. An example of boundary terms for $d=3$ is shown in Fig.~\ref{fig_3dim_terms}. 

Two technical comments follow. First the above Hamiltonian with boundary terms is not necessarily commuting. Yet, by limiting considerations to a gauge symmetric subspace $\mathcal{H}_{1}^{sym}$, the above Hamiltonian is commuting and frustration-free. One may attach projectors onto fluxless subspaces to each $\hat{O}_{v}$ term to make the Hamiltonian commuting and frustration-free. Second one can also prepare an SPT wavefunction on $\partial \Lambda$ for a subgroup $K\subseteq G$. This also leads to well-defined gapped boundary terms upon gauging with respect to $G$. For $d=2$, this $K\subseteq G$ construction gives gapped boundaries proposed by Beigi \emph{et al}. Note that Beigi \emph{et al} did not utilize the duality between SPT phases and gauge theories. In this paper, for simplicity of discussion, we mostly consider the cases where $K=G$. 

\begin{figure}[htb!]
\centering
\includegraphics[width=0.35\linewidth]{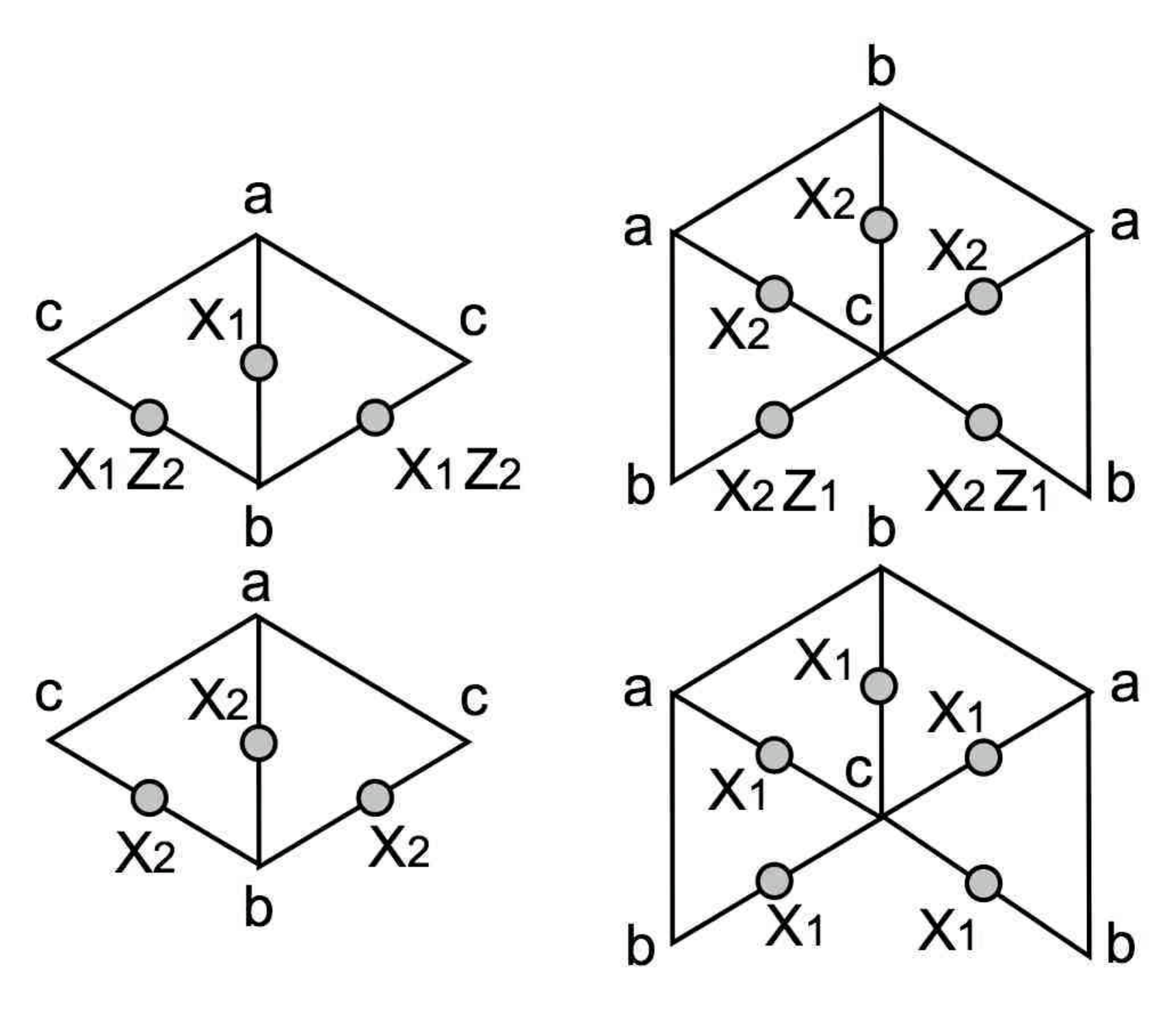}
\caption{Boundary terms for two copies of the two-dimensional toric code.
} 
\label{fig_1dim_boundary}
\end{figure}

\begin{figure}[htb!]
\centering
\includegraphics[width=0.45\linewidth]{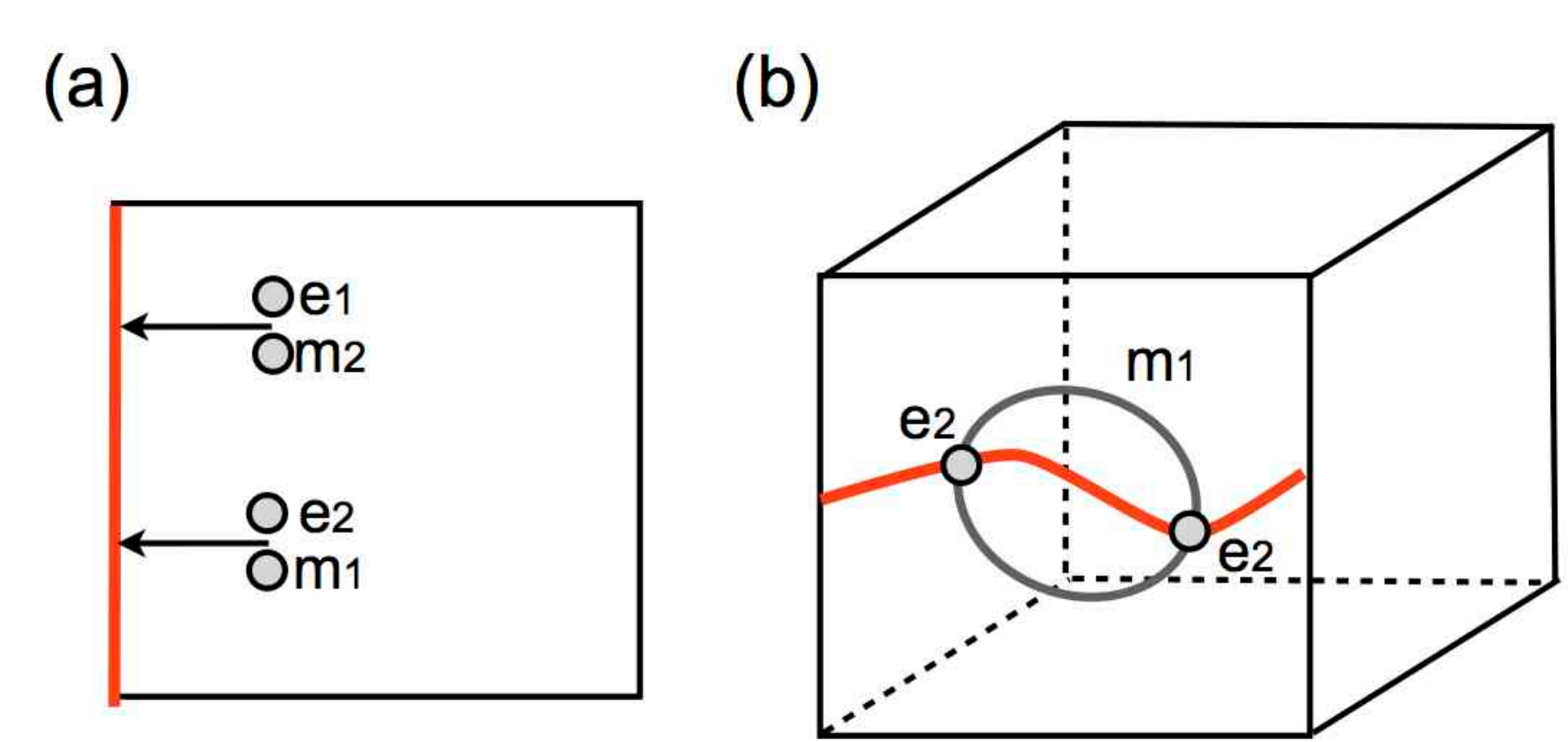}
\caption{Condensation of anyonic excitations for $G=\mathbb{Z}_{2}\otimes \mathbb{Z}_{2}$ (two copies of the toric code). (a) Two dimensions. (b) Three-dimensions. A magnetic flux and a pair of electric charges condense into the boundary. The red line represents the defect line. 
} 
\label{fig_em_boundary}
\end{figure}

It is worth looking at an example. Let $G=\mathbb{Z}_{2}\otimes \mathbb{Z}_{2}$ for $d=2$. Consider a half plane of a triangular lattice where two qubits are placed at each vertex. We shall place a non-trivial one-dimensional $\mathbb{Z}_{2}\otimes \mathbb{Z}_{2}$ SPT phase on the boundary. By gauging the Hamiltonian, one obtains two copies of the toric code with gapped boundary terms as depicted in Fig.~\ref{fig_1dim_boundary}. In two spatial dimensions, gapped boundaries of topologically ordered systems can be classified by finding a maximal set of anyonic excitations which are mutually bosonic. One may observe that $e_{1}m_{2}, e_{2}m_{1}$ can condense into the gapped boundary (Fig.~\ref{fig_em_boundary}(a)). Later we will present a generic recipe of how to find sets of condensing excitations. 

We have constructed gapped boundaries on a $(d-1)$-dimensional surface of the $d$-dimensional quantum double model by gauging a $(d-1)$-dimensional SPT wavefunction living on a $(d-1)$-dimensional surface. It is certainly possible to construct gapped boundaries by gauging $m$-dimensional SPT wavefunctions living on a $(d-1)$-dimensional surface for $m<d-1$. To be specific, consider the three-dimensional quantum double model with $G=\mathbb{Z}_{2}\otimes \mathbb{Z}_{2}$ where a gapped boundary contains a defect line corresponding to a one-dimensional non-trivial SPT wavefunction. Away from the defect line, loop-like magnetic fluxes may condense into the gapped boundary without involving any other excitations. Yet, if a magnetic flux intersects with the defect line, it cannot condense into the boundary by itself. For instance, with $G=\mathbb{Z}_{2}\otimes \mathbb{Z}_{2}$, a magnetic flux $m_{1}$ must be accompanied by a pair of electric charges $e_{2}$ near the intersections between a magnetic flux and the defect-line (Fig.~\ref{fig_em_boundary}(b)). Similarly, $m_{2}$ must be accompanied by a pair of $e_{1}$ at the intersection points.

\section{Fluctuating charges in the quantum double model}

In this section, we study $m$-dimensional fluctuating charges ($m \leq d-1$) in the $d$-dimensional quantum double model that are superpositions of point-like electric charges forming $m$-dimensional objects. We show that an $m$-dimensional fluctuating charge, living on a boundary $\partial R$ of an $(m+1)$-dimensional region $R$, can be created by a local unitary transformation acting on $R$. 

\subsection{Point-like charges and magnetic fluxes}

In the quantum double model, point-like electric charges are associated with violations of vertex terms $A_{v}$. A naturally arising question concerns the corresponding objects in a symmetric system before applying the gauging map $\Gamma$. Consider a $d$-dimensional system with global $G$-symmetry. Let us create a pair of zero-dimensional SPT phases associated with $1$-cocycles $\nu_{1}$ and $\overline{\nu_{1}}$ at vertices $v$ and $v'$ where $\overline{\nu_{1}}$ is a complex conjugate of $\nu_{1}$:
\begin{align}
|\psi\rangle = U(\nu_{1})_{v} U(\overline{\nu_{1}})_{v'} |\psi_{0}\rangle, \qquad |\psi_{0}\rangle = |+\rangle^{\otimes n}.
\end{align}
Here a symmetric local quantum circuit $U=U(\nu_{1})_{v} U(\overline{\nu_{1}})_{v'}$ is applied where $U(\nu_{1})|g\rangle = \nu_{1}(I,g)|g\rangle$. By gauging this wavefunction with respect to $G$, an output wavefunction $|\hat{\psi}\rangle$ with a pair of point-like charges at $v$ and $v'$ is obtained. For an abelian group $G$, $1$-cocycles are (linear) group representations, and there are $|G|$ different representations which correspond to different types of electric charges. Thus, for the abelian quantum double model, point-like electric charges can be associated with zero-dimensional SPT phases in a global symmetric system. 

\begin{figure}[htb!]
\centering
\includegraphics[width=0.50\linewidth]{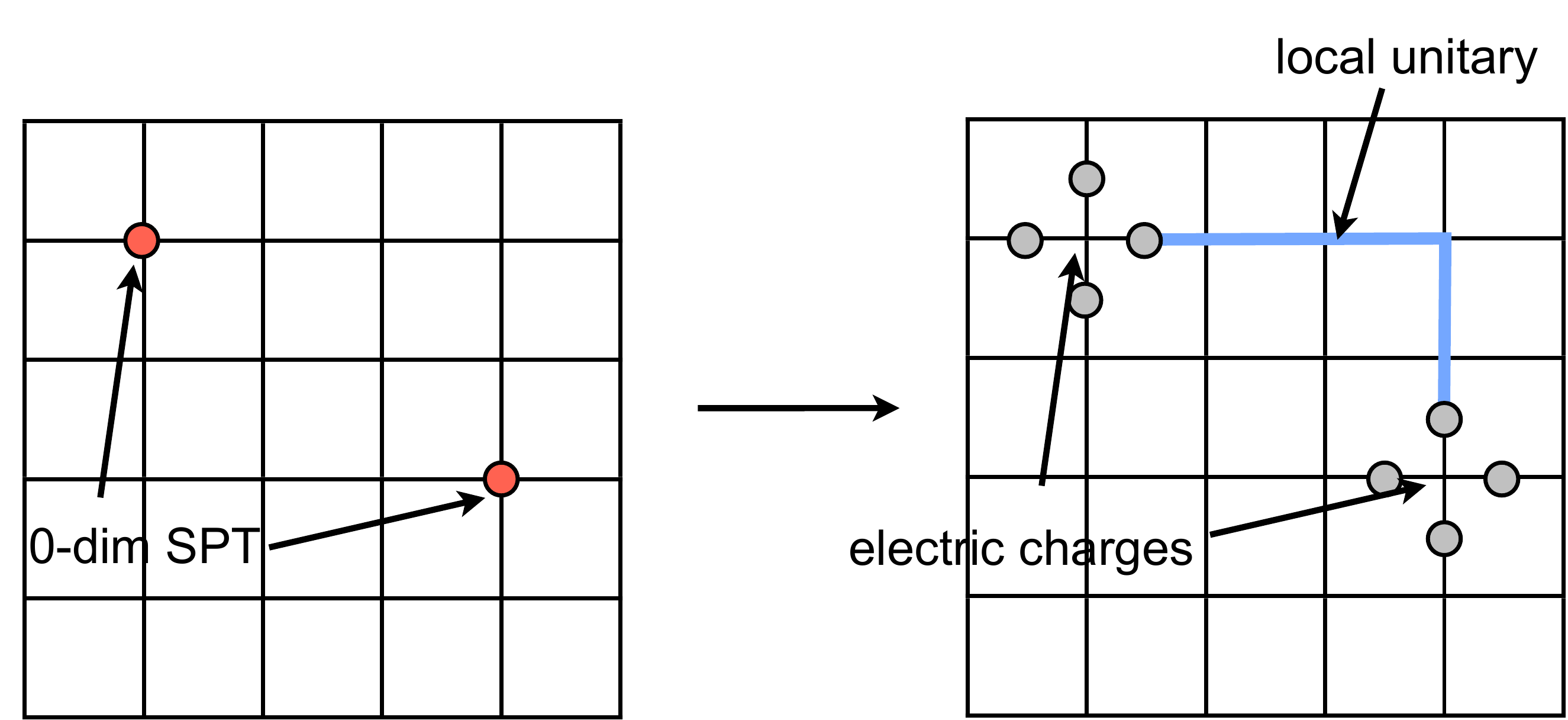}
\caption{Correspondence between $1$-cocycles (zero-dimensional SPT phases) and a pair of electric charges. 
} 
\label{fig_charge}
\end{figure}

Given a symmetric local quantum circuit $U_{v,v'}=U(\nu_{1})_{v} U(\overline{\nu_{1}})_{v'}$, there exists a corresponding local quantum circuit $\hat{U}_{v,v'}$ which creates a pair of charges at $v,v'$ in the quantum double model. Assume that $v,v'$ are neighboring vertices. Then the corresponding unitary operator $\hat{U}_{v,v'}$ is a single-body phase operator acting on an edge $e=(v,v')$. When $v,v'$ are not neighboring, consider a chain of neighboring vertices $v_{0},\ldots,v_{c}$ such that $v_{0}=v$ and $v_{c}=v'$ (Fig.~\ref{fig_charge}). The corresponding gauge symmetric circuit is given by $\hat{U}=\hat{U}_{v_{0}v_{1}}\hat{U}_{v_{1}v_{2}}\ldots \hat{U}_{v_{c-1}v_{c}}$ which is a local quantum circuit. This characterization of electric charges as zero-dimensional SPT phases works only when $G$ is an abelian group. This is due to the fact that it is not possible to move non-abelian anyons by any local quantum circuit since it would enable superluminal communication~\cite{Beckman02}.

Consider $G=\mathbb{Z}_{2}$ for $d=2$. Recall that a Pauli $Z$ operator corresponds to the following $1$-cocycle: $\nu_{1}(0,g) = (-1)^{g}$ for $g=0,1$ since $Z|g\rangle = \nu_{1}(0,g)|g\rangle$. Here an identity element is denoted by $I=0$. Consider the following symmetric wavefunction: $|\psi\rangle=Z_{v}Z_{v'}|+\rangle^{\otimes n}$. Note that $X_{v}|\psi\rangle=X_{v'}|\psi\rangle= -|\psi\rangle$. The system consists of a pair of zero-dimensional SPT phases at $v,v'$. By gauging this wavefunction, one obtains an output wavefunction $|\hat{\psi}\rangle$ which satisfies $A_{v}|\hat{\psi}\rangle=A_{v'}|\hat{\psi}\rangle=-|\hat{\psi}\rangle$, possessing electric charges at $v$ and $v'$. Here, $Z_{v}Z_{v'}$ is a symmetric local unitary, and the corresponding local unitary in a gauge theory is a string of $Z$ operators connecting $v$ and $v'$. 

\subsection{Fluctuating charges}

\begin{figure}[htb!]
\centering
\includegraphics[width=0.30\linewidth]{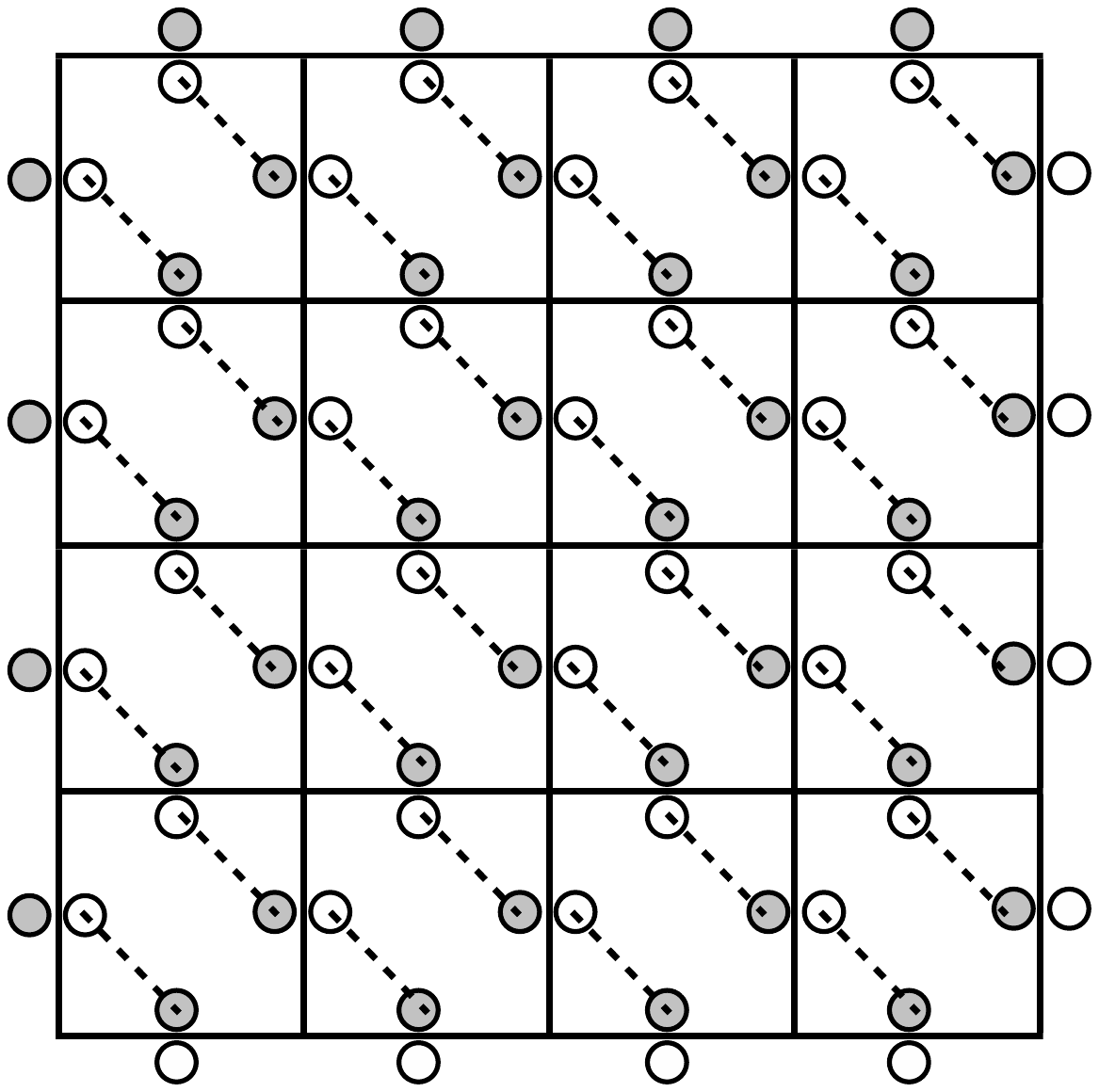}
\caption{Transversal applications of $\mbox{C}Z$ operators. Filled (white) dots represent qubits on the first (second) copy of the toric code. 
} 
\label{fig_CZ}
\end{figure}

In this subsection, we study an example of one-dimensional fluctuating charges in the two-dimensional $\mathbb{Z}_{2}\otimes \mathbb{Z}_{2}$ quantum double model. See~\cite{Beni15} for detailed discussions. Consider two decoupled copies of the toric code supported on the same lattice:
\begin{align}
H = -\sum_{v}A_{v}^{(1)}-\sum_{v}A_{v}^{(2)}-\sum_{p}B_{p}^{(1)}-\sum_{p}B_{p}^{(2)}
\end{align} 
where $A_{v}^{(i)}$ and $B_{p}^{(i)}$ are terms acting on $i$th copy of the toric code ($i=1,2$), and two qubits live on each edge of the lattice. The Control-$Z$ operator, denoted as $\mbox{C}Z$, acts on two-qubit states as follows:
\begin{align}
\mbox{C}Z|a,b\rangle = (-1)^{ab}|a,b\rangle \qquad a,b=0,1
\end{align}
which applies Pauli $Z$ operator on the second qubit if the first qubit is in the state $|1\rangle$. The control-$Z$ operator transforms Pauli operators as follows:
\begin{align}
X\otimes I \rightarrow X\otimes Z, \quad I\otimes X \rightarrow Z\otimes X,  \quad
Z\otimes I \rightarrow Z\otimes I, \quad I\otimes Z \rightarrow I\otimes Z.
\end{align}
Transversal application of $\mbox{C}Z$ operators is a logical operator since it preserves the ground state space, but has some non-trivial action on it. Namely, $\mbox{C}Z$ operators are applied on pairs of qubits in each copy of the toric code as depicted in Fig.~\ref{fig_CZ}. Let $\mbox{C}Z^{\otimes n} $ be such a transversal logical operator. Since $B_{p}^{(i)}$ consists only of Pauli $Z$ operators, $\mbox{C}Z$ operators affect vertex terms only:
\begin{align}
A_{v}^{(1)} \rightarrow A_{v}^{(1)}B_{p}^{(2)} \qquad A_{v}^{(2)} \rightarrow A_{v}^{(2)}B_{p'}^{(1)}
\end{align}
where $p,p'$ represent plaquettes that are next to the vertex $v$. Since the stabilizer group remains invariant under $\mbox{C}Z^{\otimes n}$, it is a logical operator. 

\begin{figure}[htb!]
\centering
\includegraphics[width=0.50\linewidth]{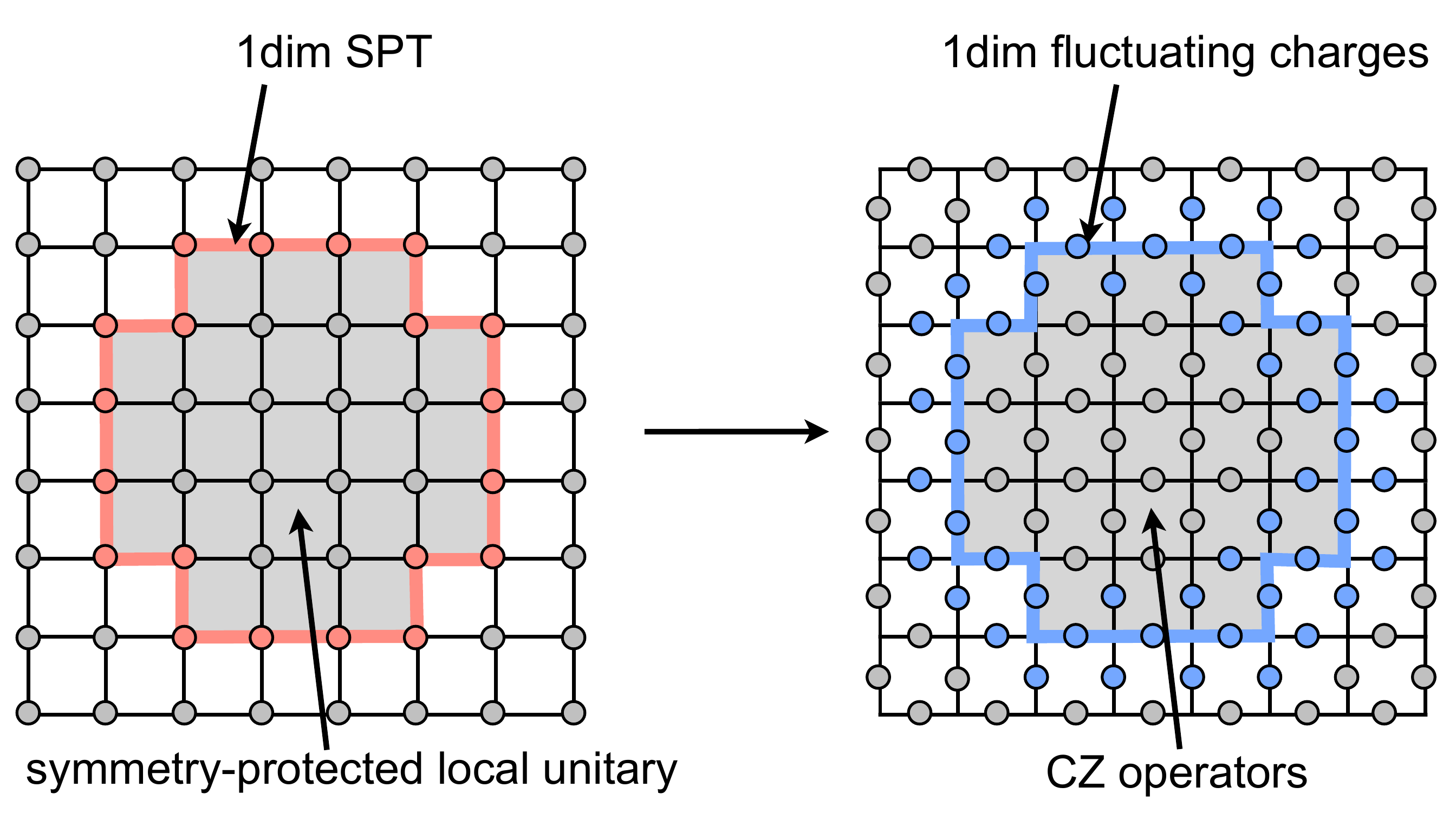}
\caption{A one-dimensional fluctuating charge from transversal $\mbox{C}Z$ operators, and the corresponding one-dimensional SPT wavefunction. 
} 
\label{fig_1dim_SPTex}
\end{figure}

\begin{figure}[htb!]
\centering
\includegraphics[width=0.70\linewidth]{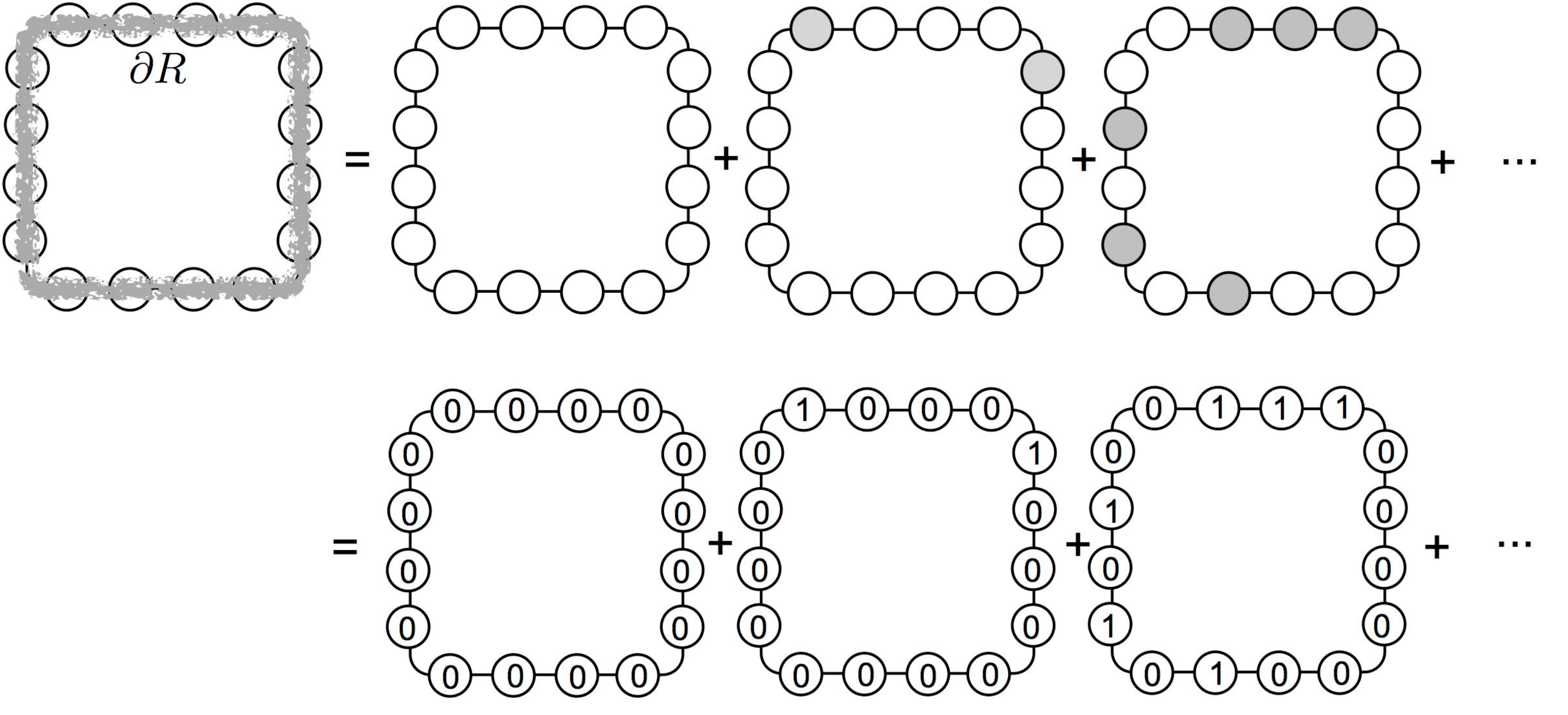}
\caption{The one-dimensional excitation, created by $\mbox{C}Z$ operators, is a superposition of point-like electric charges. Filled (white) dots represent the presence (absence) of electric charges at corresponding locations (\emph{i.e.} vertices of the square lattice). An effective one-dimensional wavefunction can be defined by assigning binary numbers $1$ and $0$, corresponding to the presence and absence of electric charges, to each dot. This one-dimensional wavefunction is identical to that of a bosonic SPT phase with $\mathbb{Z}_{2}\otimes \mathbb{Z}_{2}$ symmetries~\cite{Beni15}.
} 
\label{fig_SPT_EX}
\end{figure}

Consider a subpart of $\mbox{C}Z^{\otimes n}$ operator, denoted by ${\mbox{C}Z^{\otimes n}}_{R}$ with some contractible and connected region $R$ (Fig.~\ref{fig_1dim_SPTex}). This creates an excited state $|\psi_{ex}\rangle={\mbox{C}Z^{\otimes n}}_{R} |\psi_{gs}\rangle$ which involves excitations along the boundary $\partial R$ of $R$. Since $\mbox{C}Z^{\otimes n}$ is a diagonal phase operator, excitations will consist only of electric charges, associated with violations of $A_{v}^{(1)}$ and $A_{v}^{(2)}$. By writing the emerging wavefunction as a superposition of excited eigenstates, we find that its effective one-dimensional expression is identical to a fixed-point wavefunction of a non-trivial SPT phase with $\mathbb{Z}_{2}\otimes \mathbb{Z}_{2}$ symmetries (see Fig.~\ref{fig_SPT_EX}). The on-site $\mathbb{Z}_{2}\otimes \mathbb{Z}_{2}$ symmetries in the SPT wavefunction emerge from parity conservation of electric charges in two copies of the toric code. 

It is important to note that this one-dimensional fluctuating charge is an unbreakable loop. To create a one-dimensional fluctuating charge living on $\partial R$ where $R$ is an open disk, local unitary transformations acting on $R$ (or $\overline{R}$) are needed. Namely, there is no local unitary which creates the fluctuating charge by acting only on spins localized near $\partial R$. One can characterize this one-dimensional fluctuating charge via the gauging map. Let us consider a trivial two-dimensional symmetric system with $G=\mathbb{Z}_{2}\otimes \mathbb{Z}_{2}$, and place a non-trivial one-dimensional SPT wavefunction for $G=\mathbb{Z}_{2}\otimes \mathbb{Z}_{2}$ as shown in Fig.~\ref{fig_1dim_SPTex}. Let us gauge the system with respect to $G=\mathbb{Z}_{2}\otimes \mathbb{Z}_{2}$. Then the output wavefunction is indeed identical to the excited state which is created by transversal applications of $\mbox{C}Z$ operators. 

\subsection{SPT phases and path-integral interpretation}

Having seen an example of fluctuating charges, let us discuss their generic construction. In the $d$-dimensional quantum double model, $m$-dimensional fluctuating charges, characterized by $m$-dimensional SPT wavefunctions ($m\leq d-1$), can be created by gauge symmetric finite-depth quantum circuits. Namely, starting from a trivial symmetric wavefunction, we place an $m$-dimensional bosonic SPT wavefunction on top of it. By gauging the system with respect to $G$, an excited state with an $m$-dimensional fluctuating charge is obtained in the $d$-dimensional quantum double model. 

One technical subtlety is the parity assignment in constructing $m$-dimensional SPT wavefunctions in a $d$-dimensional graph. The strategy is to locally define the types of excitations and use them as references. Namely, a globally consistent parity assignment is possible by assigning parity to some $m$-dimensional $(m+1)$-colorable graphs and using them as references. Here we shall illustrate the idea for a one-dimensional subgraph in a $d$-dimensional graph. Pick an arbitrary $d$-simplex $\Delta$ with parity $+1$. This $d$-simplex contains $d(d-1)/2$ different $1$-simplexes. Let $a_{1},\ldots,a_{d+1}$ be color labels and $v_{1},\ldots,v_{d+1}$ be vertices on $\Delta$. We assign parity $+1$ to these $1$-simplexes. Consider an arbitrary closed one-dimensional two-colorable graph $\Lambda_{0}$ with color labels $a_{i}$ and $a_{j}$. There always exist a pair of closed one-dimensional two-colorable graph $\Lambda_{1},\Lambda_{2}$ such that $\Lambda_{1}, \Lambda_{2}$ contain the $1$-simplex $v_{i}v_{j}$ on $\Delta$ and $\Lambda_{0}=\Lambda_{1} + \Lambda_{2}$. Here the summation is XOR-type operation of edges. One can assign parity to $\Lambda_{1},\Lambda_{2}$ in an alternating way, and by using these parity assignments, one can determine parity for $\Lambda_{0}$. This treatment also works for $m$-dimensional subgraphs for arbitrary $d$. 

Key properties of fluctuating charges are summarized below.

\begin{itemize}
\item The $m$-dimensional SPT wavefunction, living on a boundary $\partial R$ of some $(m+1)$-dimensional manifold $R$, can be created by a symmetric local unitary transformation $U$ acting on R. 
\item Similarly, the $m$-dimensional fluctuating charge, living on $\partial R$, can be created by a gauge symmetric local unitary transformation $\hat{U}$ acting on $R$.
\end{itemize}

\begin{figure}[htb!]
\centering
\includegraphics[width=0.60\linewidth]{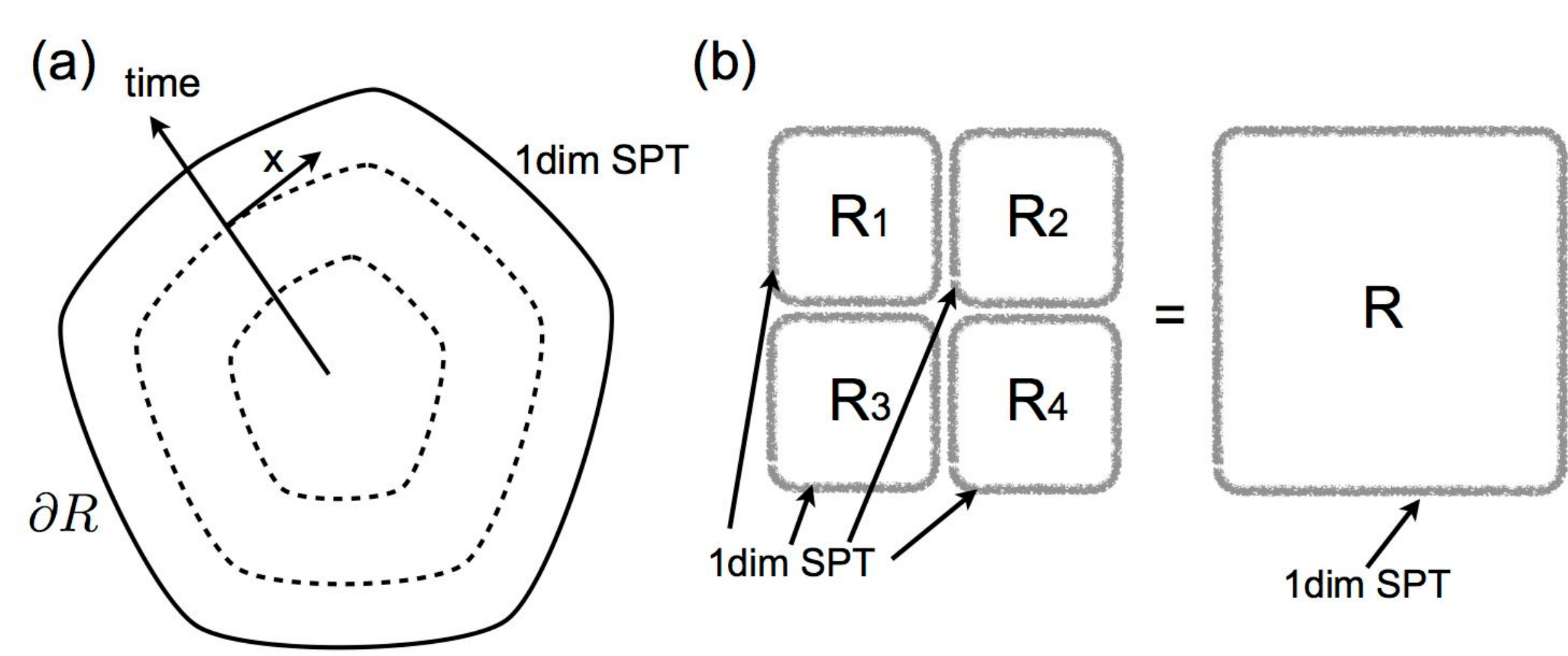}
\caption{(a) Creating a one-dimension SPT wavefunction by a symmetric local unitary in two dimensions according to the path-integral interpretation of SPT wavefunctions. (b) Attaching SPT wavefunctions to create a larger SPT wavefunction. In the language of gauge theory, this corresponds to creating a larger fluctuating charge by attaching small fluctuating charges. 
} 
\label{fig_path_integral}
\end{figure}

In order to prove these statements, it is convenient to employ the path-integral interpretation of SPT wavefunctions (Fig.~\ref{fig_path_integral}(a))~\cite{Chen13}. Consider a $(d+1)$-dimensional sphere $R$ which is $(d+2)$ colorable with color labels $a_{0},\ldots, a_{d+1}$. Consider a $d$-dimensional boundary $\partial R$ and assume that $\partial \Lambda$ is $(d+1)$-colorable with color labels $a_{1},\ldots, a_{d+1}$. Starting from a trivial symmetric wavefunction, we apply the following symmetric finite-depth quantum circuit acting on $R$:
\begin{align}
U = \prod_{\Delta} U_{\Delta}^{S(\Delta)} \qquad 
U_{\Delta}^{S(\Delta)}|g_{0}^{\Delta},\ldots,g_{d+1}^{\Delta}\rangle =\nu_{d+1}(g_{0}^{\Delta},\ldots,g_{d+1}^{\Delta})|g_{0}^{\Delta},\ldots,g_{d+1}^{\Delta}\rangle
\end{align}
The resulting wavefunction is
\begin{align}
|\psi \rangle = |\psi_{trivial}\rangle\otimes |\psi_{SPT}(\nu_{d+1})\rangle
\end{align}
where $|\psi_{SPT}(\nu_{d+1})\rangle$ is an SPT wavefunction, supported on $\partial R$, associated with the $(d+1)$-cocycle $\nu_{d+1}$ while $|\psi_{trivial}\rangle$ is a trivial symmetric wavefunction living on $R \setminus \partial R$. 
Let us split $R$ into non-overlapping regions; $R = \cup_{j}R_{j}$ and let $U_{R_{j}}$ be the unitary circuit to create an SPT wavefunction on $\partial R_{j}$. The following symmetric local unitary operator:
\begin{align}
U_{R} = \prod_{j} U_{R_{j}}
\end{align}
creates an SPT wavefunction on $\partial R$. Thus, by attaching $m$-dimensional SPT wavefunctions on $\partial R_{j}$ together, one can create a larger SPT wavefunction on $\partial R$ (Fig.~\ref{fig_path_integral}(b)). Since $U_{R_{j}}$ are symmetric local unitary circuit, $U_{R}$ is also a symmetric local unitary circuit, and there exists a corresponding gauge symmetric local unitary operator $\hat{U}_{R}$. 

These $m$-dimensional fluctuating charges are genuine $m$-dimensional objects in the sense that they cannot be created by any local quantum circuit acting exclusively on $m$-dimensional regions. To create them, one needs to apply local unitary transformations on a $(m+1)$-dimensional region. In other words, they are locally unbreakable, and thus are well-defined physical objects. If one could create $m$-dimensional fluctuating charges by $m$-dimensional local quantum circuits, then one could construct a local symmetric quantum circuit to create the underlying SPT wavefunction, leading to a contradiction.
 
Instead of applying local unitary transformations on wavefunctions, one can transform the Hamiltonian by the same local unitary transformations. In this viewpoint, a $(d-1)$-dimensional fluctuating charge can be thought of as a gapped domain wall where the Hamiltonian is modified on a $(d-1)$-dimensional hyperplane. To understand the properties of gapped domain walls, it is convenient to view a gapped domain wall as a gapped boundary by folding the entire system along the $(d-1)$-dimensional hyperplane by following Beigi \emph{et al}~\cite{Beigi11}. In this folded geometry, a gapped domain wall in the quantum double model with symmetry $G$ can be viewed as a gapped boundary in the quantum double model with symmetry $G\otimes G$ which contains two copies of the quantum double model with symmetry $G$. To construct a boundary corresponding to a domain wall, one needs to place a $(d-1)$-dimensional SPT wavefunction defined for a subgroup $K=G \subseteq G\otimes G$ where $K=\{(g,g^{-1}):g\in G\}$.
 
\section{Braiding statistics and condensations of excitations}
 
In this section, we study condensations of excitations and the braiding statistics among excitations arising in the $d$-dimensional quantum double model. Namely, we find that the multi-excitation braiding statistics among magnetic fluxes and fluctuating charges can be calculated by taking the slant products sequentially. For simplicity of discussion, we shall assume that $G$ is a finite abelian group from now on. 

\subsection{Slant product}

Here, we briefly review the notion of slant products (see~\cite{Propitius95} for instance). Given an $n$-cocycle function $\nu_{n}(g_{0},g_{1},\ldots,g_{n})$, there exists a $U(1)$-valued function $\omega_{n}(g_{1},\ldots,g_{n})$ which is one-to-one correspondence to $\nu_{n}(g_{0},g_{1},\ldots,g_{n})$, defined as
\begin{align}
\omega_{n}(g_{1},\ldots,g_{n})=\nu_{n}(I,g_{1},g_{1}g_{2},\ldots,g_{1}\cdots g_{n}).
\end{align}

\begin{itemize}
\item The slant product $i_{g}$ with $g\in G$ is a map from an $n$-cocycle to an $(n-1)$-cocycle defined as $i_{g} \omega_{n} := \omega^{(g)}_{n}$ where
\begin{equation}
\begin{split}
\omega^{(g)}_{n}(g_{1},\ldots,g_{n-1})= \ &\omega_{n}(g,g_{1},\ldots,g_{n-1}) \\ &\prod_{i=1}^{n-1} \omega_{n}(g_{1},\ldots, g_{i},g,g_{i+1},\ldots,g_{n-1})^{(-1)^{i}}.
\end{split}
\end{equation}
\end{itemize}

Sequential applications of slant products can be also defined:
\begin{align}
\omega_{n} \underset{i_{g_{1}}}\longrightarrow \omega_{n}^{(g_{1})}  \underset{i_{g_{2}}}\longrightarrow  \omega_{n}^{(g_{1},g_{2})}  \longrightarrow  \cdots \longrightarrow \omega_{n}^{(g_{1},g_{2},\ldots,g_{n-1})} \underset{i_{g_{n}}}\longrightarrow \omega_{n}^{(g_{1},g_{2},\ldots,g_{n-1},g_{n})}
\end{align}
where $\omega_{n}^{(g_{1},g_{2},\ldots,g_{n-1},g_{n})}\in U(1)$. Sequential slant products are anti-symmetric. For instance,
\begin{align}
\omega_{n}^{(g_{1},g_{2})} = ({\omega_{n}^{(g_{2},g_{1})}})^{(-1)}. 
\end{align}
One can rewrite the slant product by using $\nu_{n}$ functions:
\begin{equation}
\begin{split}
\nu^{(g)}_{n}(I,g_{1},\ldots,g_{n-1})= &\ \nu_{n}(I,g,gg_{1},\ldots,gg_{n-1}) \\ &\cdot\prod_{i=1}^{n-1} \nu_{n}(I,g_{1},\ldots, g_{i},g_{i}g,g_{i}gg_{i}^{-1}g_{i+1},\ldots,g_{i}gg_{i}^{-1}g_{n-1})^{(-1)^{i}}.
\end{split}
\end{equation}

It is worth looking at an example. Consider the following $n$-cocycle for $G=(\mathbb{Z}_{2})^{\otimes n}$:
\begin{align}
\omega_{n}(g_{1},\ldots,g_{n})= \exp( i\pi g_{1}^{(1)}\cdots g_{n}^{(n)})
\end{align}
where $g_{j}=(g_{j}^{(1)},\ldots, g_{j}^{(n)})$ with $g_{j}^{(i)}=0,1$. Define $e_{j} = (\underbrace{0,\ldots,0}_{j-1},1,0,\ldots)\in G$. Taking a slant product $i_{e_{1}}$ leads to
\begin{align}
\omega_{n}^{(e_{1})}(g_{2},\ldots,g_{n})=\exp( i\pi g_{2}^{(2)}g_{3}^{(3)}\cdots g_{n}^{(n)}).
\end{align}
By taking slant products sequentially with respect to $e_{1},\ldots,e_{n}$, one has
\begin{align}
\omega_{n}^{(e_{1},e_{2},\ldots,e_{n})}=-1.
\end{align}

\subsection{Condensation of anyons in two dimensions}

In this subsection, we study condensations of anyons in the two-dimensional quantum double model. Let us begin by characterizing point-like magnetic fluxes. Magnetic fluxes are associated with violations of plaquette-like terms $B_{p}$, and can be created at endpoints of string-like operators consisting of $L_{+}^{g}$ and $L_{-}^{g}$. To construct such a world-line operator, consider a connected region $R$ of vertices and construct an operator $\prod_{v\in R} {A^{g}}_{v}$. This operator has supports only on spins living on the boundary $\partial R$ since $L_{+}^{g}$ and $L_{-}^{g}$ operators in the interior cancel with each other for abelian $G$. This loop-like operator accounts for a process of creating a pair of $g,g^{-1}$-fluxes, moving them along the loop and then annihilating them.  One has a freedom to label a magnetic flux by $g$ or $g^{-1}$, and using either convention does not affect main results since the statistical angles just get complex conjugated. Here we use the following convention:
\begin{align}
\includegraphics[height=1.0in]{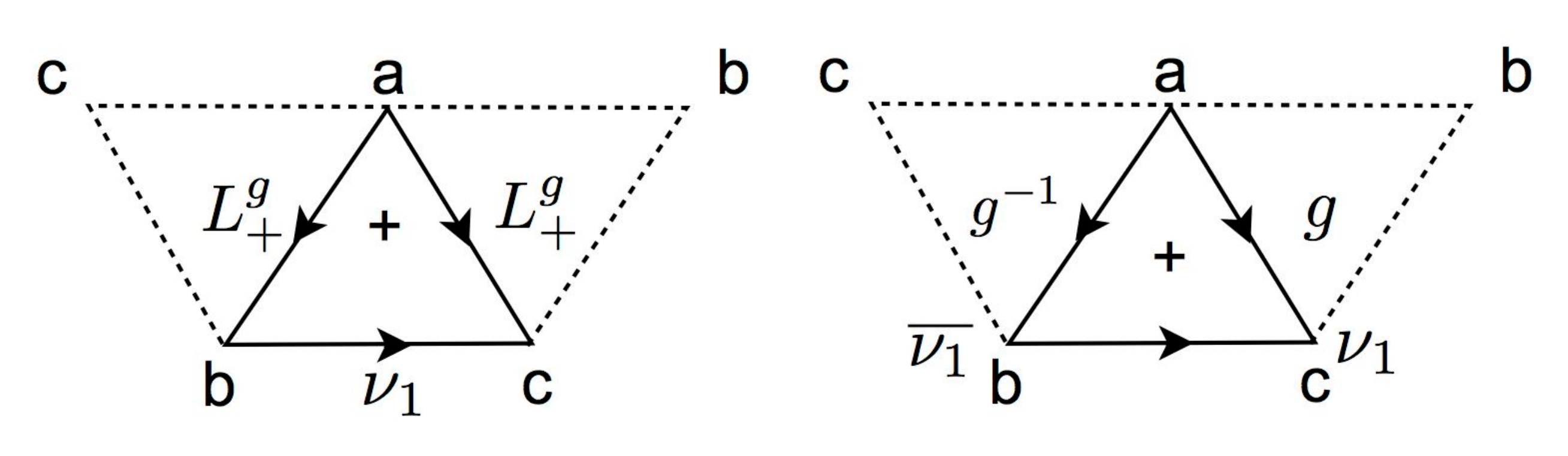}
\end{align}
In general, anyonic excitations in the two-dimensional abelian quantum double model can be labeled as a double $(g,\nu_{1})$ where $g \in G$ represents a magnetic flux and $\nu_{1}$ is a $1$-cocycle function representing the type of an electric-charge. In the above figure, $(g,\nu_{1})$ is propagating to the right. 

To find the set of condensing anyons, we will find ribbon operators that may start from and end at the gapped boundary. Consider the two-dimensional quantum double model with a gapped boundary associated with a $2$-cocycle function $\nu_{2}$. First consider the case where the $2$-cocycle $\nu_{2}$ is trivial. Let us consider $\prod_{v \in R}A_{v}$ for some connected segment $R$ on the boundary. This creates a string of $L_{+}^{g}$ and $L_{-}^{g}$ operators starting from and ending at the boundary as shown below:
\begin{align}
\includegraphics[height=1.2in]{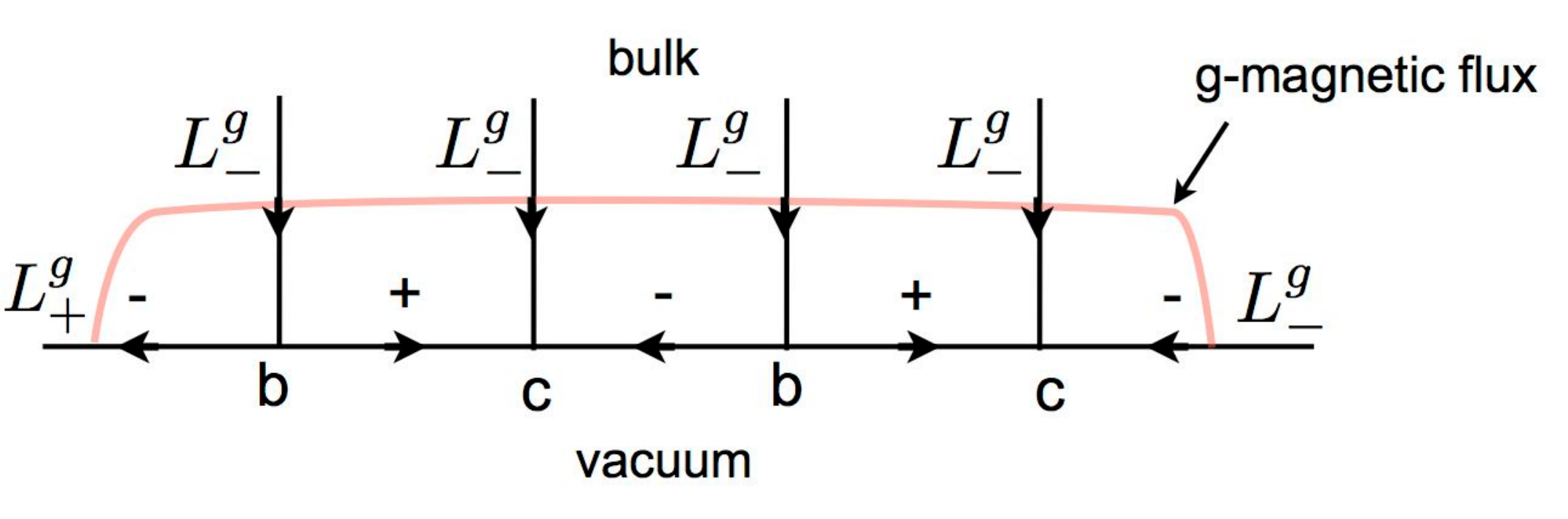}.
\end{align}
This ribbon operator accounts for a process of creating a $g$-flux from a gapped boundary, moving it through the bulk and then annihilating it on the boundary. One can deform this ribbon operator such that the anyon trajectory goes through the bulk. This implies that anyonic excitations $(g,1)$ may condense into the gapped boundary for $g\in G$.
 
\begin{figure}[htb!]
\centering
\includegraphics[width=0.40\linewidth]{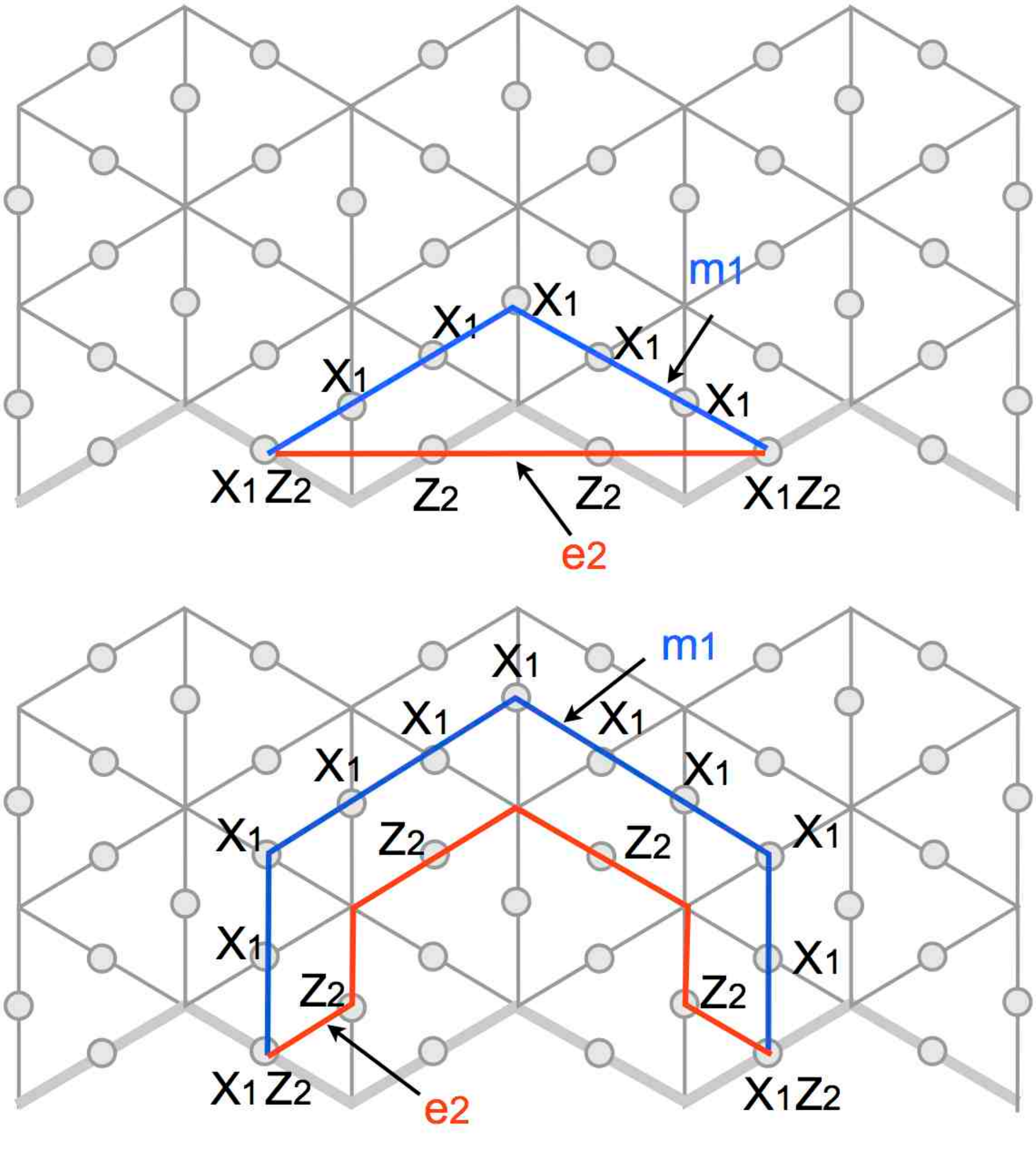}
\caption{Ribbon operators starting and ending at the boundary for $G=\mathbb{Z}_{2}\otimes \mathbb{Z}_{2}$. Ribbon operators can be deformed so that anyon trajectories go through the bulk. 
} 
\label{fig_2dim_example}
\end{figure}

When the $2$-cocycle function $\nu_{2}$ is non-trivial, vertex terms at the boundary are decorated with phase operators as shown in Fig.~\ref{fig_2dim_terms}(a)(b). By multiplying vertex operators, the following ribbon operator can be constructed:
\begin{align}
\includegraphics[height=1.25in]{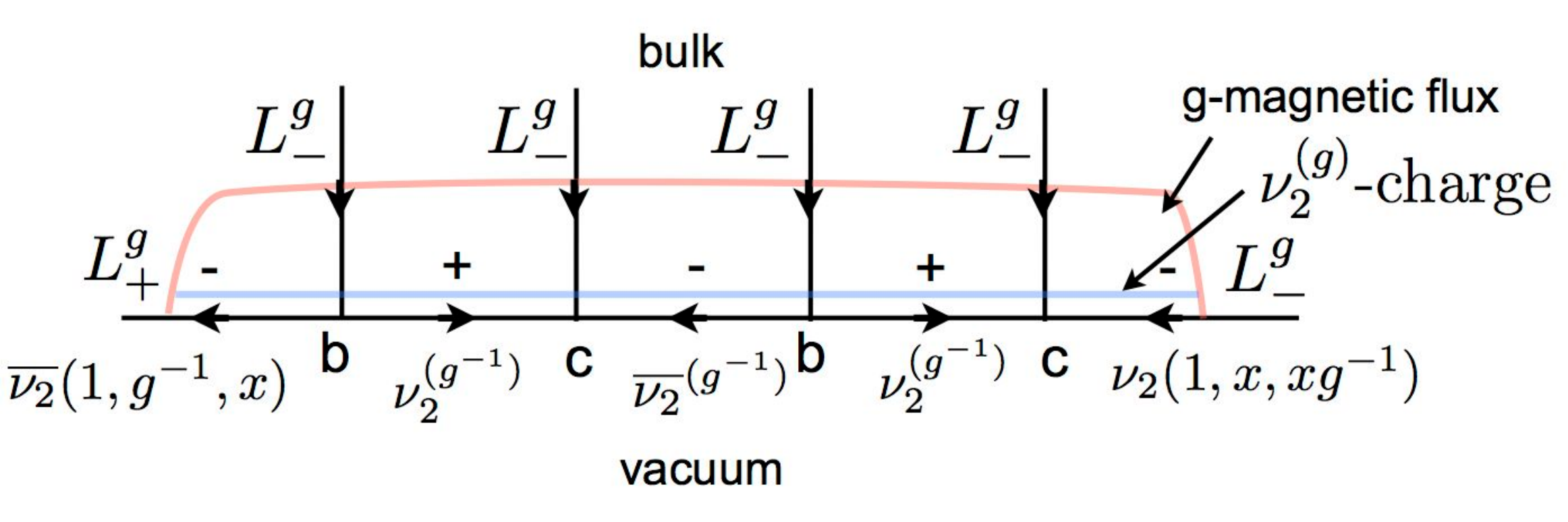}
\end{align}
where $\nu_{2}^{(g)}=i_{g}\nu_{2}$ is the slant product. This ribbon operator accounts for propagation of $E_{g}=(g,\nu_{2}^{(g)})$. Thus, $E_{g}$ may condense into the gapped boundary. When multiplying boundary terms, one needs to be careful about the fact that phase operators and $L_{+}^{g},L_{-}^{g}$ operators do not necessarily commute. Also, the assumption of $G$ being abelian is crucial. It is worth looking at an example. For $G=\mathbb{Z}_{2}\otimes \mathbb{Z}_{2}$, let us multiply boundary terms to obtain ribbon operators. Then one obtains the ribbon operator as shown in Fig.~\ref{fig_2dim_example}. This ribbon operator accounts for a process of creating a pair of $m_{1}$ and $e_{2}$ from the gapped boundary, dragging them on the bulk, and eliminating them on the boundary. 

\begin{figure}[htb!]
\centering
\includegraphics[width=0.45\linewidth]{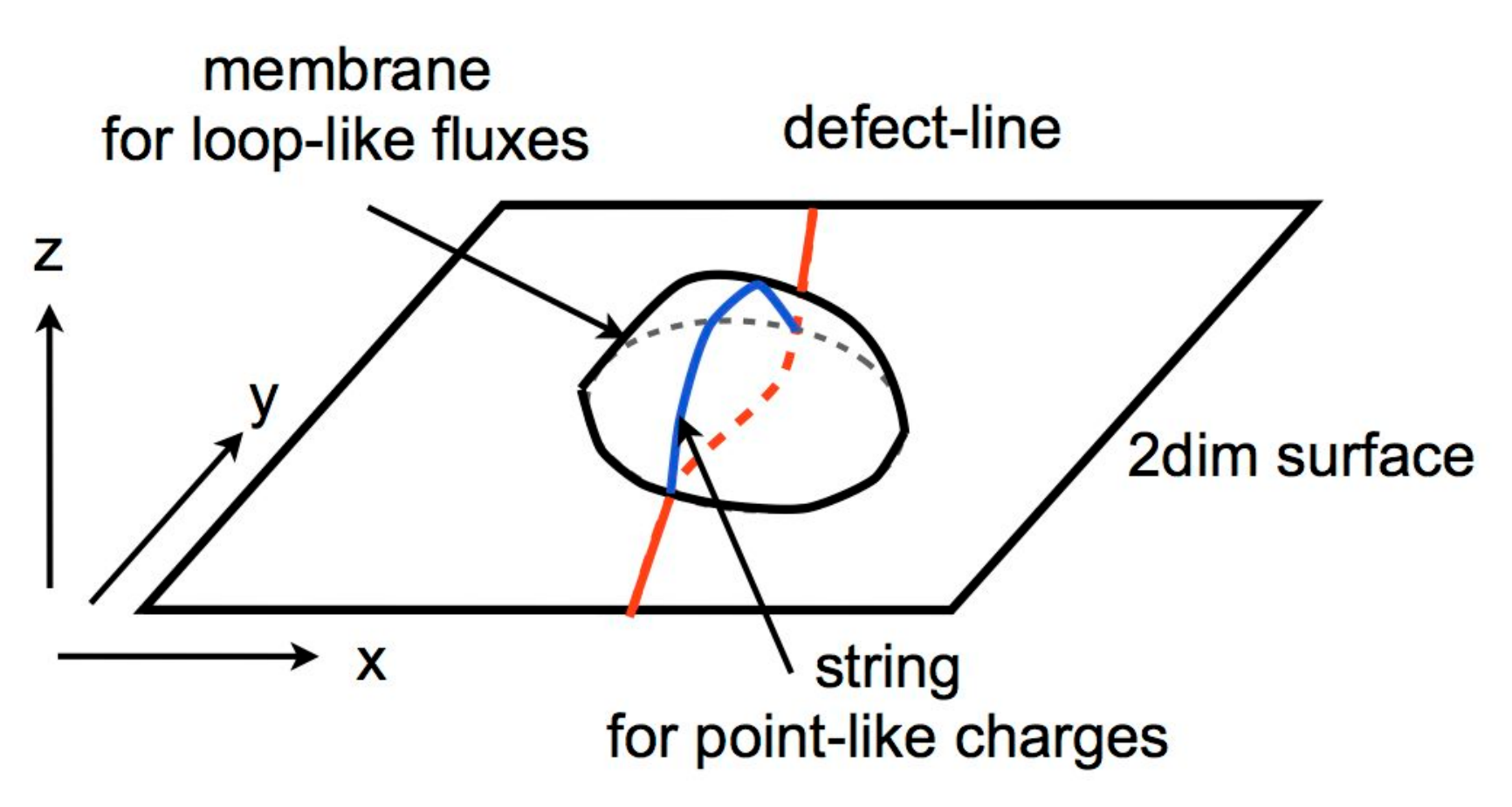}
\caption{Condensation of anyonic excitations into a boundary with a defect line in three dimensions. One can construct an operator, consisting of a world-sheet of loop-like $g$-type magnetic fluxes and a world-line of point-like $\nu_{2}^{(g)}$-type electric charges, which is anchored on the boundary. The red line represents the defect line and the blue line represents a world-line operator for an electric charge.  
} 
\label{fig_mixed_boundary}
\end{figure}

Similar characterization works for gapped boundaries with defect lines for the higher-dimensional quantum double model. Consider a defect-line associated with a $2$-cocycle function $\nu_{2}$ on a $(d-1)$-dimensional surface of the $d$-dimensional quantum double model. Then a codimension-$2$ magnetic flux may condense into a gapped boundary if it is accompanied by a pair of $\nu_{2}^{(g)}$ and $\overline{\nu_{2}}^{(g)}$ electric charges living at the intersection points with a defect-line. To see this, let us multiply boundary operators on some connected region $R$ on the $(d-1)$-dimensional surface. This leads to an operator which consists of codimension-$1$ $L^{g}$-type operators and a string-like $\nu_{2}^{(g)}$ operators. This operator characterizes propagations of $g$-type magnetic fluxes and $\nu_{2}^{(g)}$-type electric charges (Fig.~\ref{fig_mixed_boundary}). 

Let us then study the braiding statistics between magnetic fluxes and electric charges in the two-dimensional quantum double model. The braiding process can be characterized as a group commutator of ribbon operators. A group commutator between two unitary operators $A,B$ is defined as $K(A,B)=A^{\dagger}B^{\dagger}AB$. Let $A_{(g,\nu_{1})}$ be a ribbon operator corresponding to a process of creating a pair of anyons $(g,\nu_{1})$ and $(g^{-1},\overline{\nu_{1}})$ and moving $(g,\nu_{1})$ in the horizontal direction. Let $B_{(g',\nu_{1}')}$ be a similar operator which moves $(g',\nu_{1}')$ in the vertical direction. The braiding statistics between $\alpha=(g,\nu_{1})$ and $\beta=(g',\nu'_{1})$ is given by the following group commutator:
\begin{align}
K( A_{\alpha}, B_{\beta}) = e^{i\theta(\alpha,\beta)}\in U(1).
\end{align}
Note that $\theta(\alpha,\beta)=\theta(\beta,\alpha)$ since braiding $\alpha$ around $\beta$ clockwise is the same as braiding $\beta$ around $\alpha$ clockwise. To be more precise, we shall define the two-particle braiding by locally setting the following convention:
\begin{align}
\includegraphics[height=1.25in]{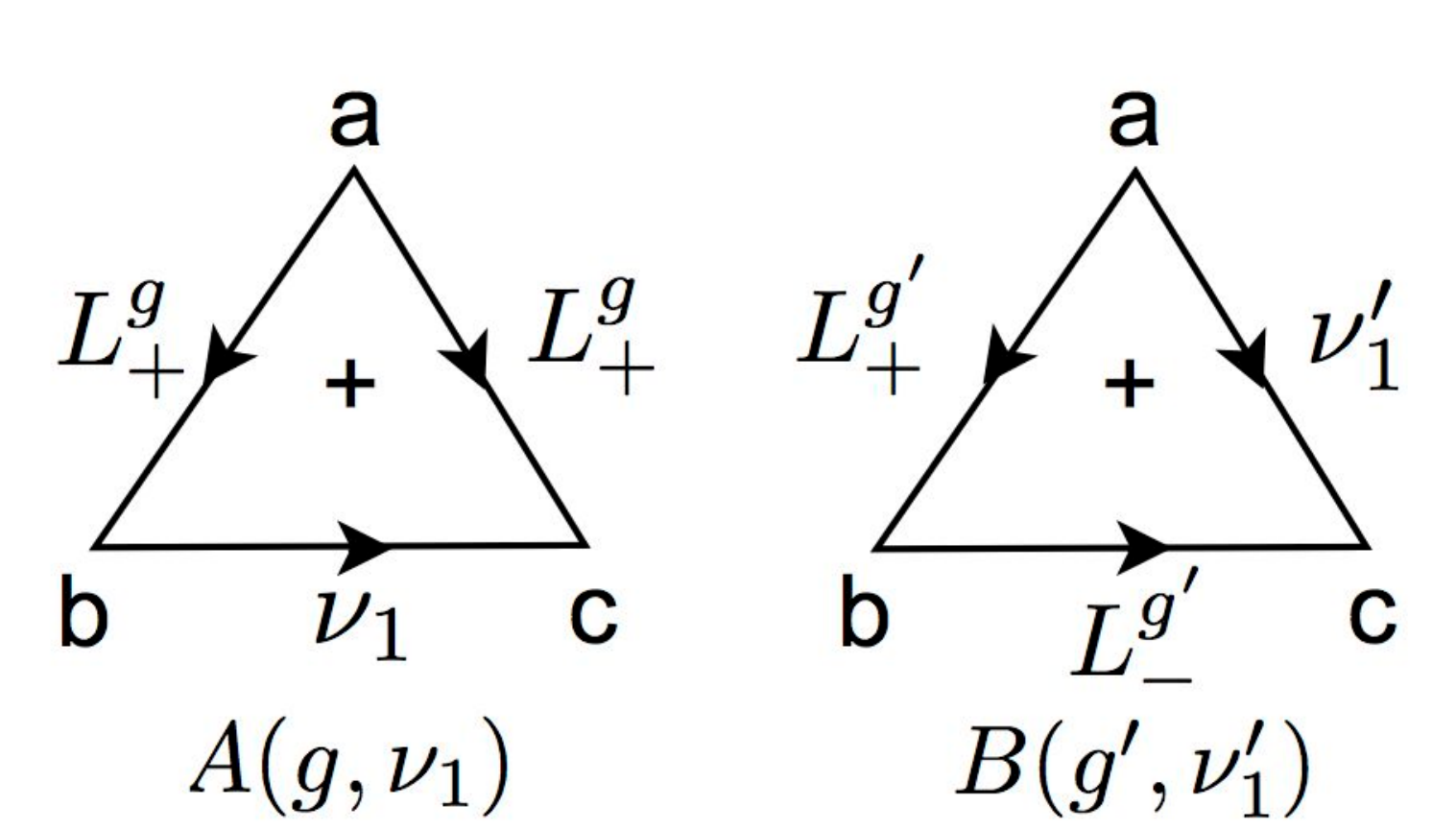}.
\end{align}
Then the statistical angle is given by
\begin{align}
e^{i\theta(\alpha,\beta)}=K(L^{g}_{+},T_{\nu_{1}'})K(T_{\nu_{1}},L^{g'}_{-})=\nu_{1}'^{(g)}\nu_{1}^{(g')}.
\end{align}
It is worth looking at an example. For $G=\mathbb{Z}_{2}$, string-operators for electric charges and magnetic fluxes are Pauli $Z$ and $X$ operators respectively, so the braiding statistics between $e$ and $m$ is $-1$. Let $0,1$ be elements of $\mathbb{Z}_{2}$. An electric charge is associated with a $1$-cocycle function $\nu_{1}(0,g)=(-1)^{g}$. Thus the braiding statistics between $m$ and $e$ is given by $\nu_{1}^{(1)}=\nu_{1}(0,1)=-1$.  

Finally, we study the braiding statistics among anyonic excitations $E_{g}$ that may condense into gapped boundaries. As for the self-statistics, observe that $(g,1)$ and $(I,\nu_{2}^{(g)})$ are mutually bosonic since $\nu_{2}^{(g,g)}=1$. So $E_{g}$ is a self-boson. As for the mutual statistics, the statistical angle between $\alpha=(g,\nu_{2}^{(g)})$ and $\beta=(h,\nu_{2}^{(h)})$ is given by
\begin{align}
\nu_{2}^{(g,h)}\nu_{2}^{(h,g)}=1.
\end{align}
This comes from the anti-symmetry of slant products. This verifies that the set of condensing anyons corresponds to the Lagrangian subgroup of anyonic excitations. From the construction of boundary ribbon operators, one can also verify that trajectories of condensing anyons commute with each other. 

\subsection{Condensation in three dimensions}

In this subsection, we find labels of anyonic excitations that may condense into gapped boundaries in the $d$-dimensional quantum double model. To be specific, we will concentrate on three-dimensional models while similar characterization works for $d>3$. We label excitations as follows:
\begin{align}
\includegraphics[height=1.2in]{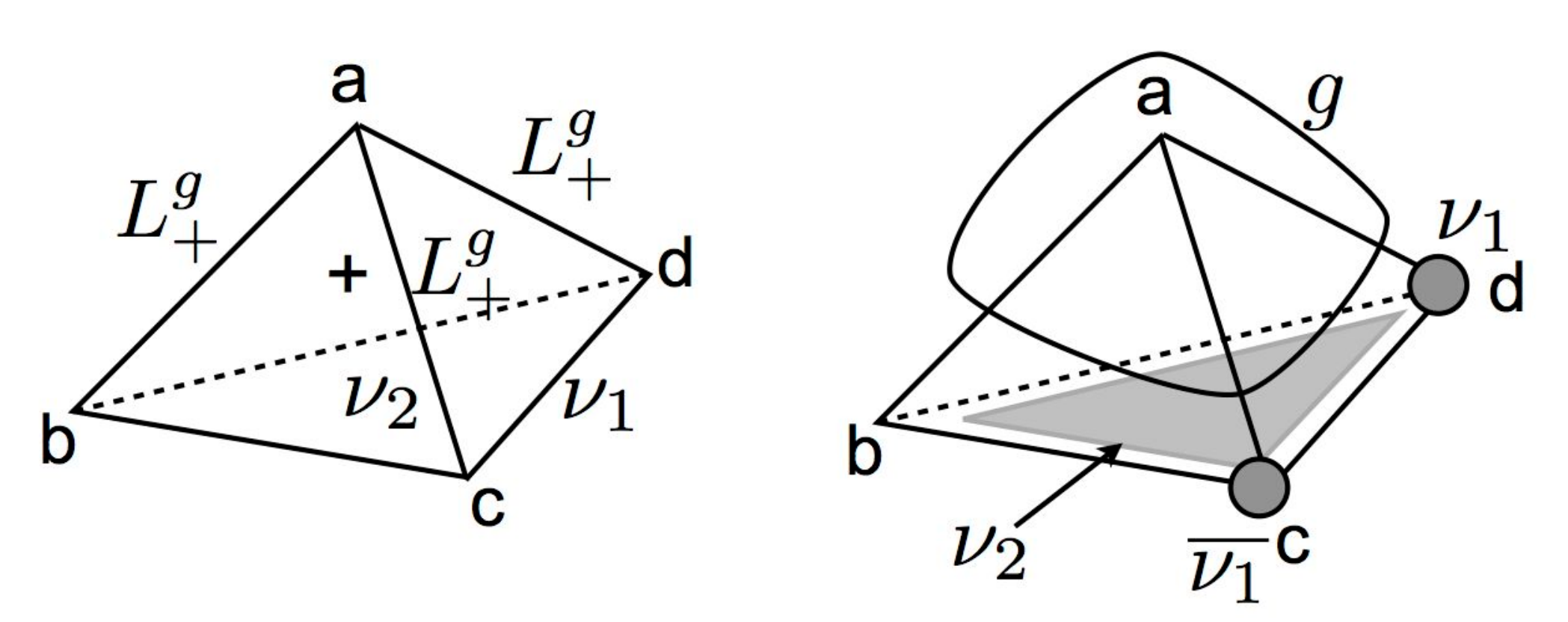}
\end{align}
which shows a $\nu_{1}$-charge, a $\nu_{2}$-charge and a $g$-flux. 

One-dimensional excitations can be labeled as a double $(g,\nu_{2})$ where $g\in G$ represents a loop-like magnetic flux and $\nu_{2}$ represents a loop-like fluctuating charge associated with an SPT wavefunction with a $2$-cocycle function $\nu_{2}$. Propagations of one-dimensional excitations can be characterized by world-sheet operators. Let us first study condensation for a trivial gapped boundary. Consider a connected region $R$ on the boundary and construct an operator $\prod_{v\in R}{A^{g}}_{v}$. This world-sheet operator has anchors supported on $\partial R$, and accounts for a process of creating/annihilating a magnetic $g$-flux at the boundary. Next, let us consider non-trivial gapped boundaries associated with $\nu_{3}$. Vertex operators on the boundary are decorated with phase operators:
\begin{align}
\hat{O}_{v^{(b)}}=A_{v^{(b)}} M_{v^{(b)}}\qquad
\hat{O}_{v^{(c)}}=A_{v^{(c)}} M_{v^{(c)}}\qquad
\hat{O}_{v^{(d)}}=A_{v^{(d)}} M_{v^{(d)}}
\end{align}
where $A_{v^{(b)}},A_{v^{(c)}},A_{v^{(d)}}$ are ordinary vertex operators while $M_{v^{(b)}},M_{v^{(c)}},M_{v^{(d)}}$ are phase operators. Here $v^{(b)},v^{(c)},v^{(d)}$ represent vertices of color $b,c,d$ respectively.  Phase operators are products of two-body phase operators as graphically shown below:
\begin{align}
\includegraphics[height=2.3in]{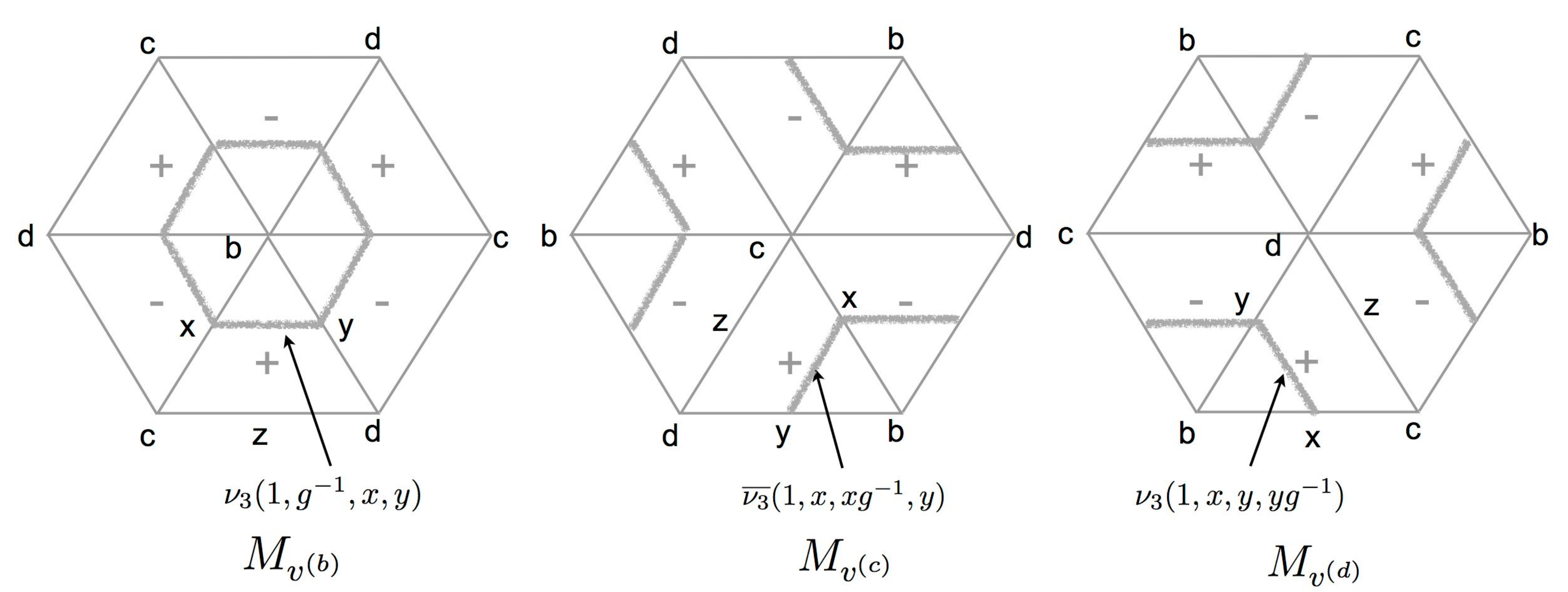}.
\end{align}  
By multiplying vertex operators on $R$, one can construct world-sheet operators which correspond to propagations of magnetic flux $g\in G$ and one-dimensional fluctuating charge $\nu_{3}^{(g)}=i_{g}\nu_{3}$. Thus, $E_{g}=(g,\nu_{3}^{(g)})$ may condense into the gapped boundary. 

\begin{figure}[htb!]
\centering
\includegraphics[width=0.35\linewidth]{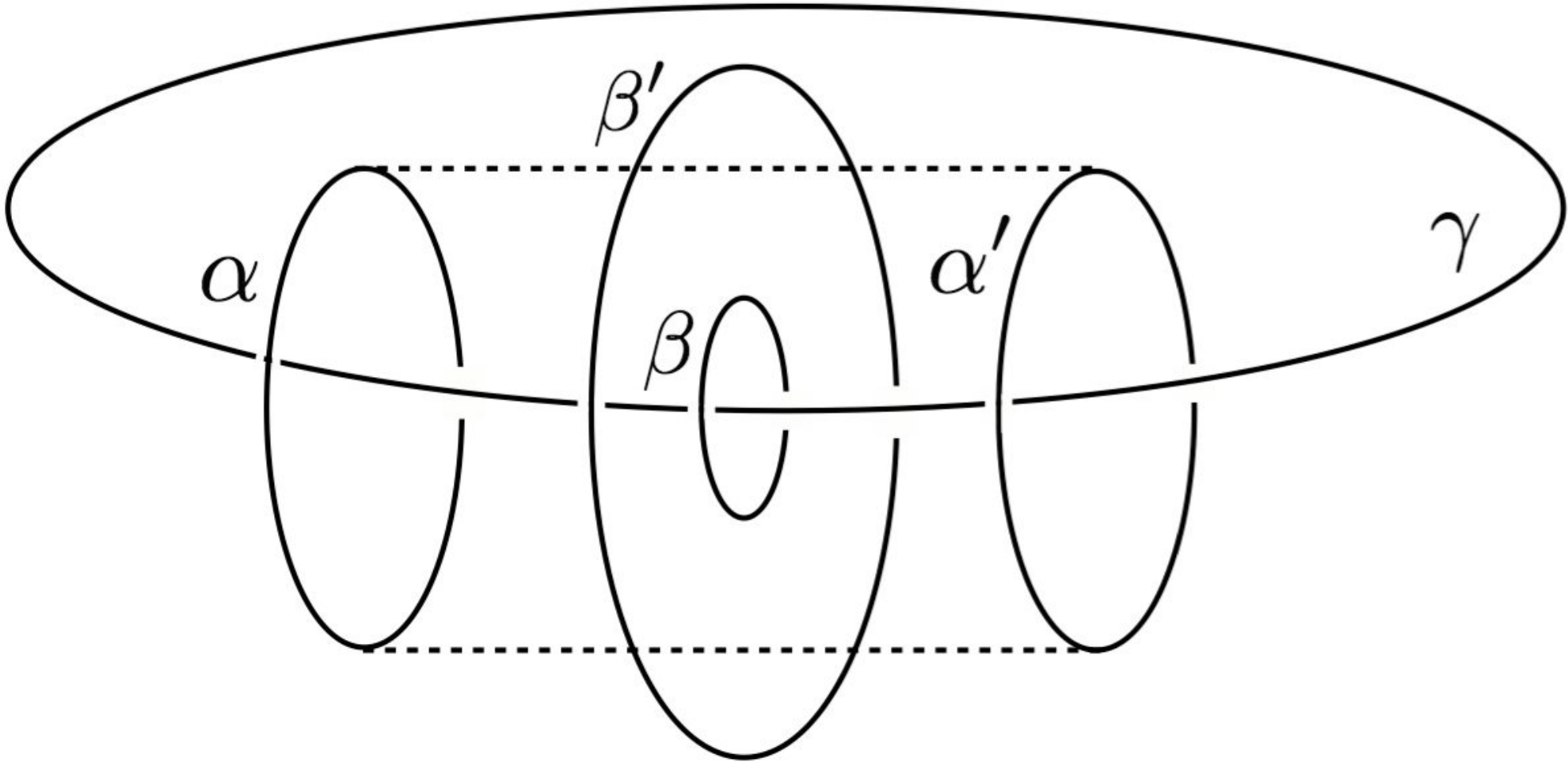}
\caption{Three-loop braiding process as a sequential group commutator $K(K(U_{\alpha}, U_{\beta}),U_{\gamma})$. 
} 
\label{fig_three_braiding}
\end{figure}

Let us then study the three-loop braiding process as depicted in Fig.~\ref{fig_three_braiding} where one first creates a loop $\gamma$, and then creates/braids $\alpha$ and $\beta$, and then brings the system back to the vacuum~\cite{Wang14, Wang15, Lin15}. The key distinction from two-loop braiding is that $\alpha$ and $\beta$ are braided in the presence of another loop $\gamma$. The braiding statistic of three loops $\alpha$,$\beta$,$\gamma$ can be found by considering a sequential group commutator. Let $U_{\alpha}$ be a world-sheet operator corresponding to creating a pair of loops $\alpha$ and $\alpha^{-1}$. Similar operators $U_{\beta},U_{\gamma}$ can be defined. Then, the following sequence of unitary transformations implements the three-loop braiding:
\begin{align}
K(K(U_{\alpha}, U_{\beta}),U_{\gamma})|\psi_{gs}\rangle = e^{i\theta(\alpha,\beta,\gamma)}|\psi_{gs}\rangle,
\end{align}
where 
\begin{align}
K(K(U_{\alpha}, U_{\beta}),U_{\gamma}) = (U_{\alpha}^{\dagger}U_{\beta}^{\dagger}U_{\alpha}U_{\beta})^{\dagger}U_{\gamma}^{\dagger}(U_{\alpha}^{\dagger}U_{\beta}^{\dagger}U_{\alpha}U_{\beta})U_{\gamma}. 
\end{align}
As such, the three-loop braiding statistics corresponds to the vacuum expectation value of the \emph{sequential} group commutator.

\begin{figure}[htb!]
\centering
\includegraphics[width=0.50\linewidth]{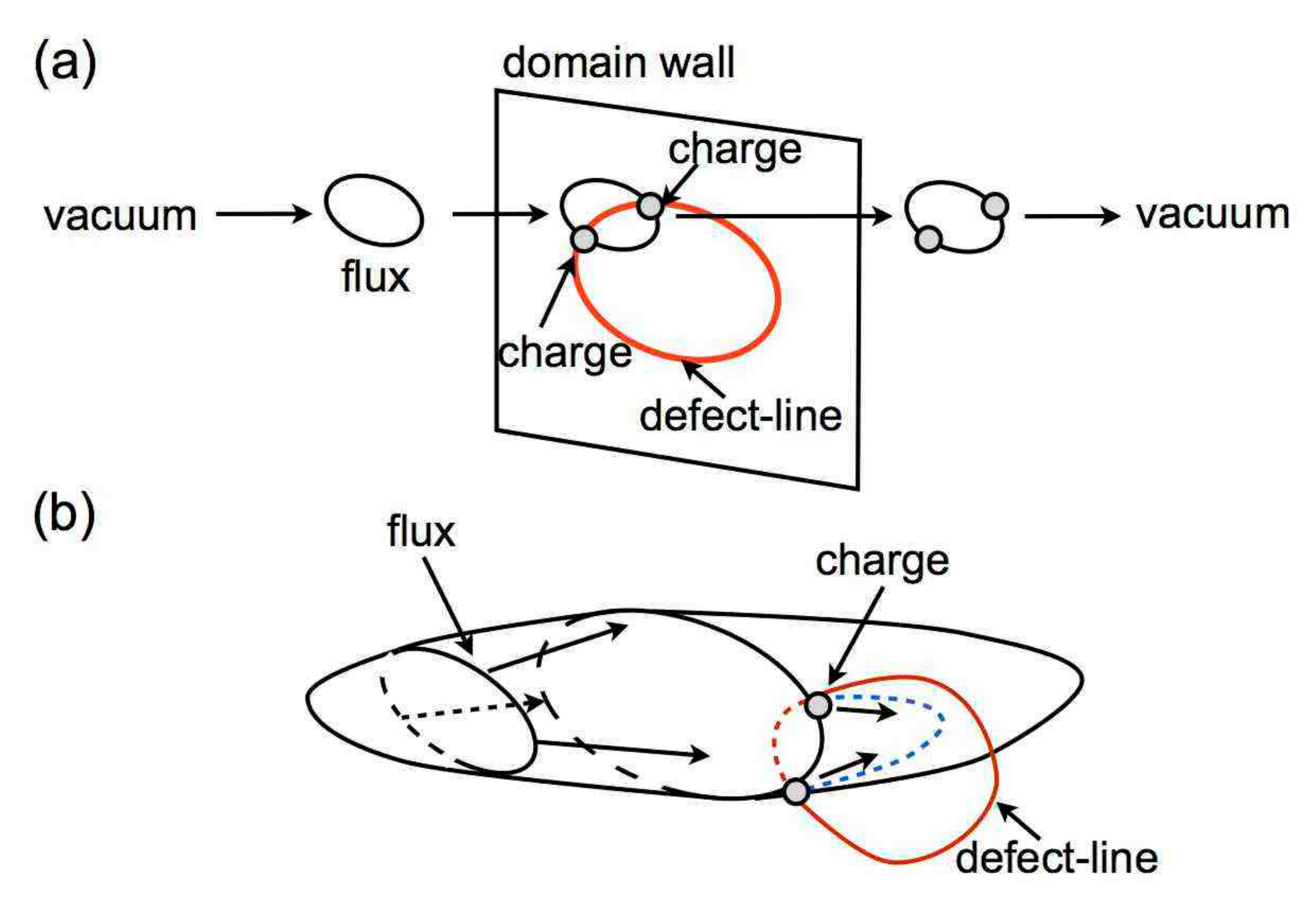}
\caption{(a) Sending a loop-like magnetic flux through a fluctuating charge. At intersection points, a pair of electric charges are created. (b) The world-sheet operator characterizing the entire process. The world-line operator, shown in blue, for electric charges are attached from the defect line, shown in red. 
} 
\label{fig_process}
\end{figure}

Directly calculating the three-loop braiding statistics is rather challenging. Yet, the statistical angle can be found from the following observation. In the three-dimensional quantum double model, consider a two-dimensional plane that contains $\beta=\nu_{2}$ loop-like fluctuating charge. One can view this plane as a gapped domain wall containing a $\nu_{2}$ defect line. Let us think of creating $\alpha=g$ magnetic flux from the vacuum and send it through the domain wall such that $\alpha$ intersects with $\beta$ (Fig.~\ref{fig_process}(a)). Upon crossing the defect line, $\alpha$ picks up a pair of electric charges characterized by $\nu^{(g)}_{2}$ and $\overline{\nu_{2}}^{(g)}$. It is convenient to consider the world-sheet operator for the entire process as shown in Fig.~\ref{fig_process}(b) where the membrane-like operator for $g$-flux picks up a world-line operator for $\nu^{(g)}$-charges which starts and ends at the intersection points with the defect line. 

\begin{figure}[htb!]
\centering
\includegraphics[width=0.50\linewidth]{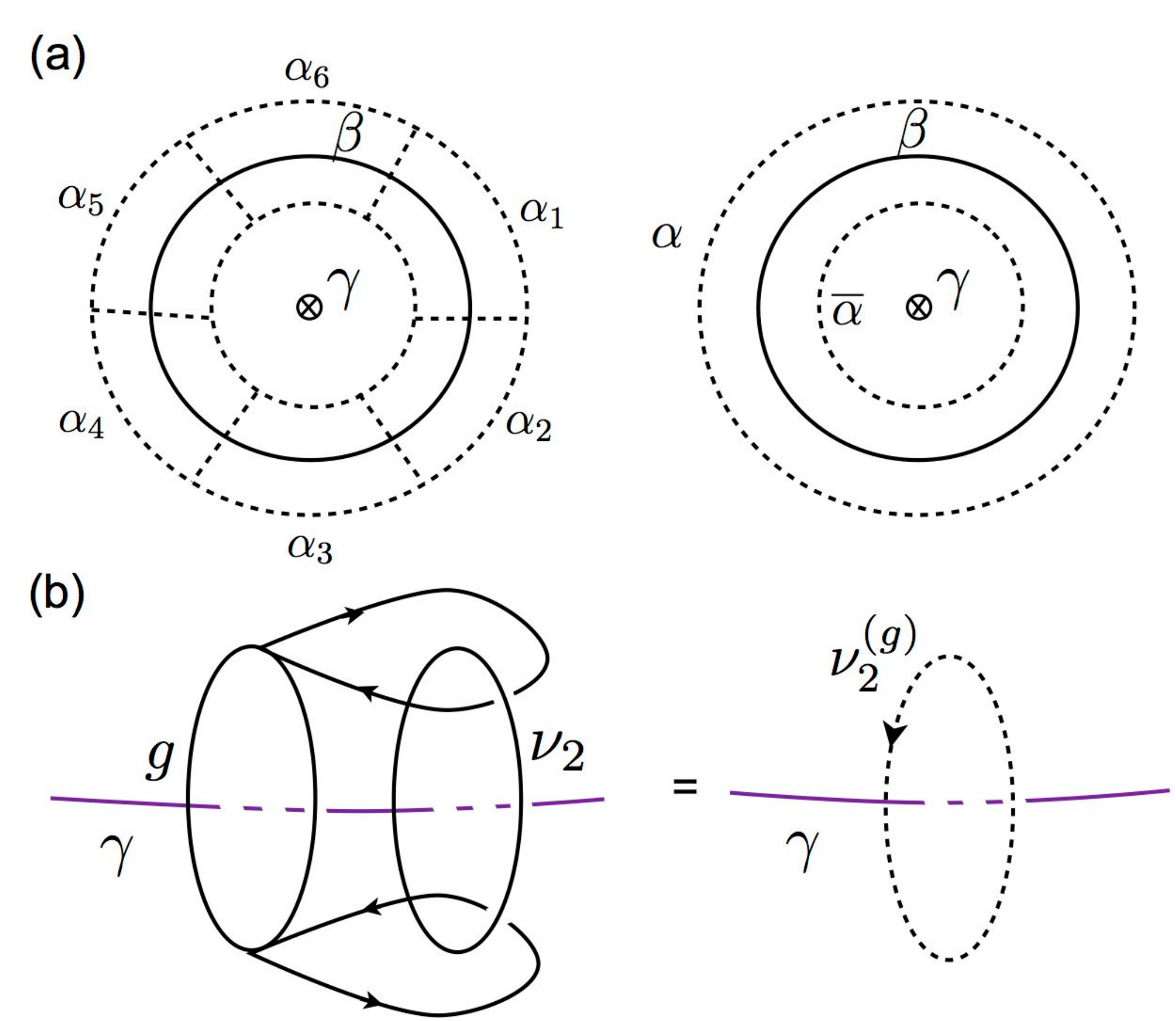}
\caption{(a) Sending multiple magnetic fluxes $\alpha_{j}$ across $\beta$. The figure depicts a plane where $\beta$ lives. Here $\alpha_{j}$ cross $\beta$ from the direction perpendicular to this plane. This is equivalent to the two-loop braiding process between $\alpha$ and $\beta$. (b) A two-loop braiding between a $g$-flux and a $\nu_{2}$-charge. As a result, a point-like $\nu_{2}^{(g)}$ charge winds around the $\gamma$-loop.  
} 
\label{fig_process2}
\end{figure}

Let us now consider a two-loop braiding process between a magnetic flux $\alpha=g$ and a fluctuating charge $\beta=\nu_{2}$ in the presence of the third loop $\gamma$ where one sends a $g$-flux from the exterior of $\nu_{2}$ and then retrieve the $g$-flux from the interior of $\nu_{2}$. This process is equivalent to sending $g$ and $g^{-1}$ in the exterior and interior of $\nu_{2}$, and then annihilate them. Let us now consider loop-like $g$-type magnetic fluxes $\alpha_{j}$ which are neighboring to each other as depicted in Fig.~\ref{fig_process2}(a) where each of magnetic fluxes $\alpha_{j}$ intersect with $\nu_{2}$. Sending $\alpha_{j}$'s across $\nu_{2}$ and then eliminating them is equivalent to sending a $g$-flux and a $g^{-1}$-flux and then eliminating them. The key observation is that, upon crossing $\nu_{2}$, world-sheet operators of $\alpha_{j}$ pick up world-line operators for $\nu^{(g)}_{2}$ electric charges (Fig.~\ref{fig_process2}(b)). By attaching these world-line operators for all the $\alpha_{j}$'s, one obtains a closed string-like world-line operator for propagation of $\nu^{(g)}_{2}$ electric charge. In other words, as a consequence of the two-loop braiding, an electric charge $\nu^{(g)}_{2}$ winds around $\gamma$. In the absence of the third loop $\gamma$, this world-line operator is contractible, and thus the braiding processes is trivial. Yet, if $\gamma$ is a magnetic flux, then the world-line operator is not contractible in general. Thus the three-loop braiding process becomes non-trivial. 

From the above observation, one finds
\begin{align}
K(U_{(g,1)},U_{(1,\nu_{2})}) = U_{\nu_{2}^{(g)}}
\end{align}
where $U_{\nu_{2}^{(g)}}$ is a closed ribbon operator for propagations of point-like charge $\nu_{2}^{(g)}$. Thus the three-loop braiding statistics is given by 
\begin{align}
K(U_{\nu_{2}^{(g)}},U_{(h,1)})=\nu_{2}^{(g,h)}.
\end{align}
In general, for the three-loop braiding process among $\alpha=(g,\nu_{2}),\beta=(g',\nu_{2}'),\gamma=(g'',\nu_{2}'')$, the statistical angle is given by
\begin{align}
e^{i\theta(\alpha,\beta,\gamma)}=\nu_{2}^{(g',g'')}\nu_{2}'^{(g,g'')}.
\end{align}
Note that $\theta(\alpha,\beta,\gamma)=\theta(\beta,\alpha,\gamma)$. Let us then study braiding statistics among condensing excitations $E_{g}=(g,\nu_{3}^{(g)})$. The statistical angle for the three-loop braiding among $E_{g_{1}},E_{g_{2}},E_{g_{3}}$ is given by
\begin{align}
\nu_{3}^{(g_{1},g_{2},g_{3})}\nu_{3}^{(g_{2},g_{1},g_{3})}=1.
\end{align}
This implies that the set of condensing excitations possess mutually trivial three-loop braiding statistics. 

It is worth looking at an example. Consider the three-dimensional quantum double model with $G=\mathbb{Z}_{2}\otimes \mathbb{Z}_{2} \otimes \mathbb{Z}_{2}$. Note that the system is equivalent to three decoupled copies of the three-dimensional toric code, which is also known as the three-dimensional topological color code in quantum information community. Let us consider the boundary associated with a $3$-cocycle function $\omega_{3}(g_{1},g_{2},g_{3})=(-1)^{g_{1}^{(1)}g_{2}^{(2)}g_{3}^{(3)}}$ where $g_{j}=(g_{j}^{(1)},g_{j}^{(2)},g_{j}^{(3)})$. The non-trivial slant products are
\begin{align}
\omega_{3}^{([1,0,0])}(g_{1},g_{2})=(-1)^{g_{1}^{(2)}g_{2}^{(3)}}\qquad 
\omega_{3}^{([0,1,0])}(g_{1},g_{2})=(-1)^{g_{1}^{(1)}g_{2}^{(3)}}\qquad 
\omega_{3}^{([0,0,1])}(g_{1},g_{2})=(-1)^{g_{1}^{(1)}g_{2}^{(2)}}.
\end{align}
So, the following excitations may condense into the gapped boundary
\begin{align}
(m_{1},s_{23}),(m_{2},s_{13}),(m_{3},s_{12}).
\end{align}
Here $m_{1},m_{2},m_{3}$ are loop-like magnetic fluxes from each copy of the toric code. Also $s_{12}$ is a loop-like fluctuating charge consisting of charges from the first and the second copies of the toric code. The three-loop braiding statistics is given by 
\begin{align}
e^{\theta(s_{ij},m_{i},m_{j})}=e^{\theta(m_{i},s_{ij},m_{j})}=-1 \qquad i\not=j.
\end{align}
From this, we can verify that condensing excitations exhibit mutually trivial three-loop braiding statistics. This three-loop braiding statistics has been studied in~\cite{Beni15}.

\section{Fault-tolerantly implementable logical gate}

In this section, we comment on applications of our construction to the problem of finding/classifying fault-tolerantly implementable logical gates in topological quantum codes. 

\subsection{Fault-tolerant logical gate and the Clifford hierarchy}

One possible route to fault-tolerant quantum computation is to perform computational tasks inside a protected subspace (codeword space) of a quantum error-correcting code. An important question concerns how to implement logical gates inside the codeword space. Ideally, one hopes to perform logical gates by a finite-depth local quantum circuit since, if a logical gate implementation requires a highly non-local and complicated quantum circuit, local errors may propagate to the entire system in an uncontrolled manner. As such, it is crucial to find/classify fault-tolerantly implementable logical gates in quantum error-correcting codes~\cite{Gottesman99, Eastin09, Bravyi13b, Bombin06, Bombin15, Beverland14, Pastawski15, Kubica15, Kubica15b}~\footnote{There are logical gates which do not admit finite-depth circuit implementation, but can be implemented in a rather simple manner. For instance, a Hadamard-like logical gate can be implemented in the two-dimensional toric code by shifting the lattice sites in a diagonal direction, followed by transversal applications of Hadamard operators~\cite{Wen03}. Also, a lattice rotation may implement a topological $S$-matrix. However, in this paper, for simplicity of discussion, we call logical gates which can be implemented by a finite-depth quantum circuit fault-tolerantly implementable logical gates.}. Despite the importance of the problem, no systematic procedure of finding fault-tolerant  logical gates is known. 

In~\cite{Bravyi13b}, Bravyi and K{\"o}nig derived a bound on the power of fault-tolerantly implementable logical gates in topological stabilizer codes. To state their result, let us begin by recalling the notion of the Clifford hierarchy. Consider a system of $n$ qubits, and the set of all the Pauli operators, denoted by $\mathsf{Pauli} = \langle X_j, Y_j, Z_j \rangle_{j \in [ 1, n]}$. We define the \emph{Clifford hierarchy} as $\mathcal{C}_0 \equiv  U(1)$ (\emph{i.e.} global complex phases), and then recursively as
  \begin{equation}
    \mathcal{C_{}}_{m + 1} = \{ U : \forall P \in \mathsf{Pauli},\  K(P,U)
    \in \mathcal{C}_m \}.
  \end{equation}
where $K(P,U)=P^{\dagger}U^{\dagger}PU$ is a group commutator. By definition, the sizes of sets $\mathcal{C}_{m}$ are increasing: $\mathcal{C}_{m}\subset\mathcal{C}_{m+1}$. Note that $\mathcal{C}_{m}$ is a set and is not a group for $m\geq 3$. At the lowest level, $\mathcal{C}_{1}$ is a group of Pauli operators with global complex phases while $\mathcal{C}_{2}$ coincides with the so-called \emph{Clifford group}. The Gottesman-Knill theorem tells that any quantum circuit composed exclusively from Clifford gates in $\mathcal{C}_2$, with computational basis preparation and measurement, may be efficiently simulated by a classical computer~\cite{Nielsen_Chuang}. In contrast, incorporating any additional non-Clifford gate to $\mathcal{C}_2$ results in a universal gate set. 

Bravyi and K{\"o}nig considered the set of logical gates that may be realized on a topological stabilizer code via a finite-depth local quantum circuit. By topological stabilizer  codes, we mean that the code generators (interaction terms in the Hamiltonian) are supported on geometrically local regions. Their main result is stated as follows.

\begin{theorem}[Bravyi-K{\"o}nig]
Consider a family of stabilizer codes with geometrically local generators in $d$ spatial dimensions with a code distance growing in the number of physical qubits. Then any fault-tolerant logical unitary, fully supported on an $m$-dimensional region ($m\leq d$), has a logical action included in $\mathcal{C}_{m}$.
\end{theorem}

So, in $d$ spatial dimensions, fault-tolerant logical gates are restricted to the $d$th level of the Clifford hierarchy. It is known that the $d$-dimensional $G=(\mathbb{Z}_{2})^{\otimes d}$ quantum double model (which is a topological stabilizer code) saturates the bound by Bravyi-K{\"o}nig~\cite{Kubica15b}. Namely, one can implement a $d$-qubit control-$Z$ gate fault-tolerantly by a finite-depth quantum circuit as we will see below. The connection between group cohomology and the Clifford hierarchy can be established by considering sequential group commutators. If $U$ is a non-trivial $m$th-level operator, belonging to $\mathcal{C}_{m}$ but outside of $\mathcal{C}_{m-1}$, then there exist $m$ non-trivial Pauli operators $P_{j}$ for $j=1,\ldots, m$ such that 
\begin{align}
K(\ldots K(K(U,P_{1}),P_{2})\ldots ,P_{m})\not = 1. 
\end{align}
For instance, letting $P_{j}=X_{j}$ for a system of $m$ qubits, one has 
\begin{align}
K(\ldots K(K(\mbox{C}^{\otimes m-1}Z,X_{1}),X_{2})\ldots ,X_{m})= -1
\end{align}
where $\mbox{C}^{\otimes m-1}Z$ is the $m$-qubit control-$Z$ gate.

\subsection{Logical gates from gapped domain wall}

Let us briefly recall the connection between the notion of fault-tolerant logical gates and gapped domain walls. Consider some topological quantum error-correcting code, such as the $d$-dimensional quantum double model, and assume that there exists a non-trivial $d$-dimensional logical gate $U$ that can be implemented by a local unitary circuit. Let us split the entire system into the left and right parts and apply the logical gate $U$ only on the right hand side of the system. This transforms the Hamiltonian into the following form:
\begin{align}
H = H_{left} + H_{wall} + H_{right}
\end{align}
where $H_{left}$ and $H_{right}$ may remain unchanged while $H_{wall}$ can be viewed as a gapped domain wall which connects $H_{left}$ and $H_{right}$. Since the logical gate $U$ is non-trivial, we expect that excitations are transposed upon crossing the domain wall. This observation implies that, given a $d$-dimensional non-trivial fault-tolerant logical gate, one can construct a corresponding domain wall since non-trivial logical gates would transform types of excitations. 

\begin{figure}[htb!]
\centering
\includegraphics[width=0.55\linewidth]{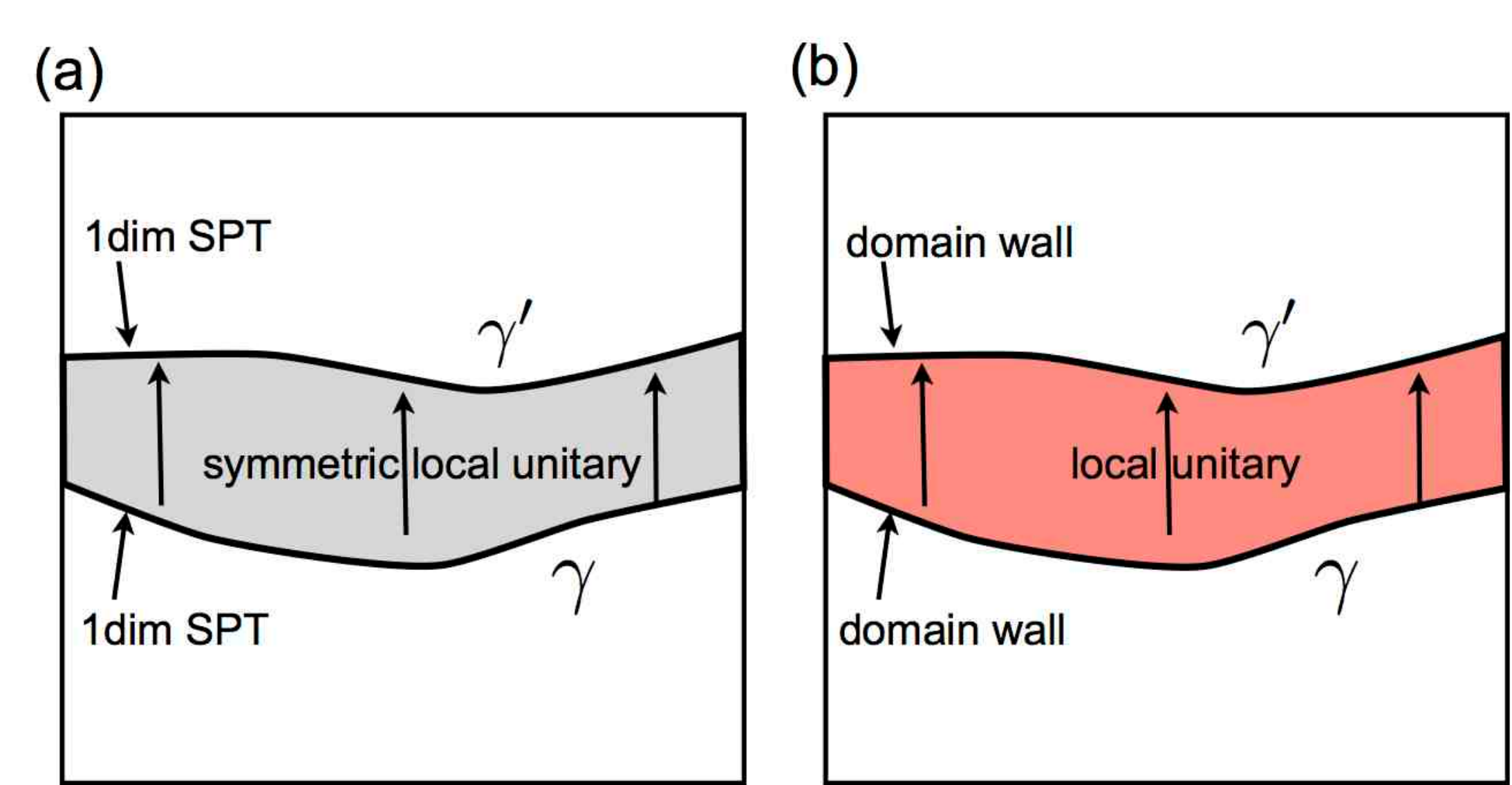}
\caption{A fault-tolerant logical gate constructed from SPT wavefunctions and gapped domain walls. (a) Moving a one-dimensional SPT wavefunction by finite-depth symmetric quantum circuits. (b) Moving a gapped domain wall by finite-depth quantum circuits.
} 
\label{fig_logical_gate}
\end{figure}

By reversing the argument, a fault-tolerant logical gate may be constructed from a gapped domain wall. Namely, consider a $(d-1)$-dimensional SPT wavefunction and the corresponding domain wall in the gauge theory (Fig.~\ref{fig_logical_gate}(a)(b)). Assume that the associated $d$-cocycle function has a non-trivial sequence of slant products. Note that one can sweep the domain wall over the entire system by a finite depth circuit. This operation will have non-trivial action on the codeword space. Thus, one can construct a $d$-dimensional fault-tolerantly implementable logical gate by using $d$-cocycle functions over $G$ in the $d$-dimensional quantum double model.

Let us apply the above idea to the quantum double model with $G=\mathbb{Z}_{2}^{\otimes m}$ which is supported on a $d$-torus. Since the system is a topological stabilizer code, the result by Bravyi and K{\"o}nig holds. The system has $2^{k}$ degenerate ground states with $k=md$. The system has anti-commuting pairs of one-dimensional $Z$-type logical operators and codimension-$1$ $X$-type logical operators, which can be denoted as 
\begin{align}
\overline{X}^{g}_{\ell},\overline{Z}^{g}_{\ell},\qquad g\in G, \qquad \ell = 1,\ldots, d
\end{align}
where $\overline{Z}^{g}_{\ell}$ extends in the $\hat{\ell}$ direction while $\overline{X}^{g}_{\ell}$ is perpendicular to the $\hat{\ell}$ direction. Ground states can be labelled as 
\begin{align}
|\psi\rangle = |g_{1}\rangle\otimes |g_{2}\rangle \otimes \ldots \otimes |g_{d}\rangle,\qquad g_{j}\in \mathbb{Z}_{2}^{\otimes m}
\end{align}
where $\overline{X}^{h}_{\ell}|\psi\rangle =|g_{1}\rangle\otimes \ldots \otimes | g_{\ell}+h\rangle \otimes \ldots \otimes |g_{d}\rangle$. Here we choose $|I\rangle \otimes |I\rangle \otimes \ldots \otimes |I\rangle$ to be the ground state with no global flux. Let $U_{\nu_{d}}$ be a finite depth circuit which sweeps the corresponding domain wall, associated with a $d$-cocycle function $\nu_{d}$, over the entire system. Sequential group commutators are given by
\begin{align}
K(K(U_{\nu_{d}},\overline{X}^{g_{1}}_{1}),\overline{X}^{g_{2}}_{2})\ldots , \overline{X}^{g_{d}}_{d}) = \nu_{d}^{(g_{1},\ldots,g_{d})}.\label{eq:logical_commutator}
\end{align}
This is because one can view the above sequential commutator as a braiding process among $(d-1)$-dimensional excitations in $(d+1)$ dimensions. Thus, if the sequential slant product is non-trivial, then the logical unitary $U_{\nu_{d}}$ is a non-trivial $d$th-level Clifford gate. By using Eq.~(\ref{eq:logical_commutator}) to $|I\rangle \otimes |I\rangle \otimes \ldots \otimes |I\rangle$, one finds
\begin{align}
U_{\nu_{d}} |g_{1}\rangle\otimes |g_{2}\rangle \otimes \ldots \otimes |g_{d}\rangle = \nu_{d}^{(g_{1},g_{2},\ldots,g_{d})}|g_{1}\rangle\otimes |g_{2}\rangle \otimes \ldots \otimes |g_{d}\rangle\label{eq:permutation}.
\end{align}
It is worth looking at an example. For $G=\mathbb{Z}_{2}^{\otimes d}$, consider a non-trivial $d$-cocycle $\omega_{d}(g_{1},\ldots,g_{d})=(-1)^{g_{1}^{(1)}\ldots g_{d}^{(d)}}$ which has a non-trivial sequence of slant products. The corresponding logical gate has the following action:
\begin{align}
U_{\nu_{d}} |g_{1}\rangle\otimes |g_{2}\rangle \otimes \ldots \otimes |g_{d}\rangle = \prod_{(a_{1},\ldots,a_{d})} (-1)^{g_{1}^{(a_{1})}\ldots g_{d}^{(a_{d})} }|g_{1}\rangle\otimes |g_{2}\rangle \otimes \ldots \otimes |g_{d}\rangle
\end{align}
where $(a_{1},\ldots,a_{d})$ is a permutation of $(1,2,\ldots,d)$. This logical operator corresponds to a transversal application of $R_{d}$ phase operators in the $d$-dimensional topological color code. The logical action resembles that of the $d$-qubit control-$Z$ gate. 

Finally, we extend the discussion to arbitrary abelian $G$ and generalize the Clifford hierarchy. For a single $|G|$-dimensional spin, Pauli operators are defined as
\begin{align}
L_{+}^{g} = \sum_{h\in G} |gh\rangle \langle h | \qquad 
T_{\nu_{1}}= \sum_{h\in G } \nu_{1}(1,h)|h\rangle \langle h|.
\end{align}
For a system of $n$ spins, one can define the generalized Clifford hierarchy by taking $\mathcal{C}_{1}$ to be the Pauli operator group generated by $L_{+}^{g}$ and $T_{\nu_{1}}$. Then one can recursively define the generalized hierarchy $\mathcal{C}_{m}$. A straightforward generalization of Bravyi-K{\"o}nig argument implies that logical gates which can be implemented by a finite depth circuit are restricted to $\mathcal{C}_{d}$ in the $d$-dimensional quantum double model. By labeling ground states as $|\psi\rangle = |g_{1}\rangle\otimes |g_{2}\rangle \otimes \ldots \otimes |g_{d}\rangle$ where $g_{j}\in G$ as before, Eq.~(\ref{eq:permutation}) holds for any abelian $G$. If the sequential slant product is non-trivial, then the corresponding logical gate belongs to the generalized $\mathcal{C}_{d}$, but is outside of $\mathcal{C}_{d-1}$. 

\section{Open questions}

There remain a number of interesting open problems and future questions. Some of them are listed below:

\begin{enumerate}[(a)]
\item Recently there have been discussions on SPT phases protected by $q$-form symmetry operators where charged excitations have $q$ spatial dimensions~\cite{Kapustin13, Kapustin14c, Gaiotto15, Beni15b}. It is an interesting future problem to consider their implications to gapped boundaries and fault-tolerant logical gates. 
\item This paper provides a number of interesting gapped domain walls and boundaries for higher-dimensional systems. Developing a mathematical framework (which perhaps utilizes higher-category) to classify higher-dimensional gapped boundaries is an important project. 
\item In this paper, we considered excitations made of bosonic particles. Can we create excitations characterized by other types of particles such as fermions? 
\item Spatial dimension of symmetry operators can be non-integer values~\cite{Haah11, Bravyi13, Beni13}. Namely, one may construct an SPT Hamiltonian protected by fractal-like symmetry operators. 
\item Fault-tolerant logical operators can be viewed as global symmetries of the Hamiltonian. By imposing them on the system, one may be able to explore novel symmetry-enriched topological phases of matter.
\item One can consider a unitary operator corresponding to a non-trivial $d$-cocycle function whose slant products are trivial. 
While, on a torus, such an operator is trivial, whether it is trivial or not on other geometries is an interesting question. 
\item Our result seems to suggest that, for $d$-dimensional qubit stabilizer codes, at least $d$ copies of the toric code are necessary in order to admit a non-trivial $d$th level gate for $d>2$ since, with less than $d$ copies, sequential slant products become trivial at some point. Can we prove/disprove this conjecture? 
\item Sequential slant products are important in finding the number of ground states on a torus~\cite{J_Wang15c}. The relation between ground state degeneracy and gapped boundaries may be an interesting question.
\item How do we formulate condensations, multi-excitation braiding and fluctuating charges in the language of field theories, say in topological BF theory?
\end{enumerate}

\section*{Acknowledgment}

I would like to thank Isaac Kim, Fernando Pastawski, John Preskill and Chenjie Wang for helpful discussions and/or comments. Part of the work was completed during the visits to the Kavli Institute for Theoretical Physics. We acknowledge funding provided by the Institute for Quantum Information and Matter, an NSF Physics Frontiers Center with support of the Gordon and Betty Moore Foundation (Grants No. PHY-0803371 and PHY-1125565). I was supported by the David and Ellen Lee Postdoctoral fellowship. This research was supported in part by the National Science Foundation under Grant No. NSF PHY11-25915. Research at Perimeter Institute is supported by the Government of Canada through Industry Canada and by the Province of Ontario through the Ministry of Research and Innovation.


\end{document}